\documentclass[12pt]{article} %***
\usepackage[sectionbib]{natbib}
\usepackage{array,epsfig,fancyheadings,rotating}
\usepackage[]{hyperref}  %<----modified by Ivan
\usepackage{sectsty, secdot}
\sectionfont{\fontsize{12}{14pt plus.8pt minus .6pt}\selectfont}
\renewcommand{\theequation}{\thesection\arabic{equation}}
\subsectionfont{\fontsize{12}{14pt plus.8pt minus .6pt}\selectfont}
\usepackage{amsmath}
\usepackage{amssymb}
\usepackage{amsfonts}
\usepackage{multirow}
\usepackage{amsthm}
\usepackage{setspace}
\usepackage{tikz}
\usetikzlibrary{shapes,decorations,arrows,calc,arrows.meta,fit,positioning}
\tikzset{
    -latex,auto,node distance =2 cm and 2 cm, on grid, very thick,
    state/.style ={ellipse, draw, minimum width = 1 cm},
    point/.style = {circle, draw, inner sep=0.04cm,fill,node contents={}},
    bidirected/.style={Latex-Latex,dashed},
    el/.style = {inner sep=2pt, align=left, sloped}
}

\newtheorem{theorem}{Theorem}

\newtheorem{corollary}{Corollary}
\newtheorem{proposition}{Proposition}
\theoremstyle{definition}

\newtheorem{example}{Example}

\usepackage{color}
\usepackage{multirow}
\usepackage{xcolor}
\usepackage{bm} 
\usepackage{enumitem}

\numberwithin{equation}{section}

\newcommand{\bcol}[1]{{\color{black} #1}}

\def\pr{\textup{pr}}
\newcommand{\ind}{\perp \!\!\! \perp}

\newcommand\T{{ \mathrm{\scriptscriptstyle \top} }}
\begin{document}

%%%%%%%%%%%%%%%%%%%%%%%%%%%%%%%%%%%%%%%%%%%%%%%%%%%%%%%%%%%%%%%%%%%%%%%%%%%%%%%%%%%%%%%%%%%%%%%%%%%%%%%%%%%%%%%%%%%%%%%%%%%%
%%%%%%%%%%%%%%%%%%%%%%%%%%%%%%%%%%%%%%%%%%%%%%%%%%%%%%%%%%%%%%%%%%%%%%%%%%%%%%%%%%%%%%%%%%%%%%%%%%%%%%%%%%%%%%%%%%%%%%%%%%%%

\renewcommand{\baselinestretch}{2}

\markright{ \hbox{\footnotesize\rm Statistica Sinica
%{\footnotesize\bf 24} (201?), 000-000
}\hfill\\[-13pt]
\hbox{\footnotesize\rm
%\href{http://dx.doi.org/10.5705/ss.20??.???}{doi:http://dx.doi.org/10.5705/ss.20??.???}
}\hfill }

\markboth{\hfill{\footnotesize\rm FIRSTNAME1 LASTNAME1 AND FIRSTNAME2 LASTNAME2} \hfill}
{\hfill {\footnotesize\rm FILL IN A SHORT RUNNING TITLE} \hfill}

\renewcommand{\thefootnote}{}
$\ $\par

%%%%%%%%%%%%%%%%%%%%%%%%%%%%%%%%%%%%%%%%%%%%%%%%%%%%%%%%%%%%%%%%%%%%%%%%%%%%%%%%%%%%%%%%%%%%%%%%%%%%%%%%%%%%%%%%%%%%%%%%%%%%

\fontsize{12}{14pt plus.8pt minus .6pt}\selectfont \vspace{0.8pc}
\centerline{\large\bf Identification and estimation of treatment effects on long-term}
\vspace{2pt} 
\centerline{\large\bf outcomes in clinical trials with external observational data}
\vspace{.4cm} 
\centerline{Wenjie Hu$^{a}$, Xiao-Hua Zhou$^{a, b}$ and Peng Wu$^{c\ast}$\footnote{$\ast$correspond to: pengwu@btbu.edu.cn.}}  
\vspace{.4cm} 
\centerline{$^a$Peking University, $^b$Pazhou Lab, $^c$Beijing Technology and Business University} 
\vspace{.55cm} \fontsize{9}{11.5pt plus.8pt minus.6pt}\selectfont

% long-term treatment effect
%%%%%%%%%%%%%%%%%%%%%%%%%%%%%%%%%%%%%%%%%%%%%%%%%%%%%%%%%%%%%%%%%%%%%%%%%%%%%%%%%%%%%%%%%%%%%%%%%%%%%%%%%%%%%%%%%%%%%%%%%%%%

\begin{quotation}
\noindent {\it Abstract: In biomedical studies, estimating drug effects on chronic diseases requires a long follow-up period, which is difficult to meet in randomized clinical trials (RCTs). The use of a short-term surrogate to replace the long-term outcome for assessing the drug effect relies on stringent assumptions that empirical studies often fail to satisfy.
Motivated by a kidney disease study, we investigate the drug effects on long-term outcomes by combining an RCT without observation of long-term outcomes and an observational study in which the long-term outcome is observed but unmeasured confounding may exist.  Under a mean exchangeability assumption weaker than the previous literature, 
we identify the average treatment effects in the RCT and derive the associated efficient influence function and semiparametric efficiency bound. Furthermore, we propose a locally efficient doubly robust estimator and an inverse probability weighted (IPW) estimator. The former attains the semiparametric efficiency bound if all the working models are correctly specified, which may be hard to achieve due to the intertwined working models. While the latter has a simpler form and requires much fewer model specifications. The IPW estimator using estimated propensity scores is more efficient than that using true propensity scores and achieves the semiparametric efficient bound in the case of discrete covariates and surrogates with finite support. 
 Both estimators are shown to be consistent and asymptotically normally distributed.  
 Extensive simulations are conducted to evaluate the finite-sample performance of the proposed estimators. 
We apply the proposed methods to estimate the efficacy of oral hydroxychloroquine on renal failure in a real-world data analysis.}

\vspace{9pt}
\noindent {\it Key words and phrases:}
Data fusion, Long-term treatment effects, Semiparametric efficiency, Surrogate.    \par 
\end{quotation}\par

\def\thefigure{\arabic{figure}}
\def\thetable{\arabic{table}}

\renewcommand{\theequation}{\thesection.\arabic{equation}}

\fontsize{12}{14pt plus.8pt minus .6pt}\selectfont

% =========================================================================
% Section 1

\section{Introduction}
\label{sec1}

In biomedical research,  randomized clinical trials (RCTs) are the gold standard for drug or therapy evaluation  \citep{cartwright2010randomised}.
However, the high cost of labour and material resources restricts the sample size and the duration of RCTs. 
Especially for chronic diseases, the important long-term outcomes are difficult to observe during the period of RCTs.   
As a motivating example, we consider a clinical study on immunoglobulin A  nephropathy (IgAN), which is the most prevalent form of primary glomerular disease worldwide \citep{d1987commonest}. A double-blind randomized clinical trial is conducted at Peking University First Hospital for six months, to compare the efficacy of additional use of oral hydroxychloroquine (HCQ)  with only optimized renin-angiotensin-aldosterone system (RAAS) inhibition, which is a standard therapy for IgAN disease. 
The outcome of interest is whether the patient will develop renal failure over a period of time. 
 But IgAN is a chronic disease and the long-term outcome is not observed within a six-month experiment period, instead, the researcher collected the percentage change in proteinuria as a surrogate \citep{liu2019effects}.
 
This kind of need to evaluate the long-term effect in clinical trials is pervasive in medical and social science applications and requires new methodologies. If only RCT data is available, to evaluate the effect of a treatment on the long-term outcome, the researchers often choose a short-term surrogate that is strongly predictive of the outcome and can be observed during the RCTs and then report analysis results for the surrogate \citep{liu2019effects}.   
The criteria for choosing short-term surrogates have been studied over the years \citep{prentice1989surrogate,frangakis2002principal,lauritzen2004discussion,chen2007criteria,ju2010criteria}.  
However, the aim of using a surrogate to replace the outcome of interest is too ambitious \citep{kallus2020role}. For example, \citet{chen2007criteria} raised the surrogate paradox, a phenomenon that treatment has a positive effect on a surrogate that has a positive effect on the outcome of interest, but the treatment has a negative effect on the outcome of interest. Stringent unverifiable assumptions are made to avoid the surrogate paradox \citep{chen2007criteria}. 
Thus it is important to propose more flexible methods that rely on less stringent assumptions to estimate the treatment effects on the long-term outcomes.
Besides RCT data,  hospitals usually have a large amount of observational data containing long-term outcomes. 
Nevertheless, the existence of unmeasured confounders is unavoidable in an observational study \citep{Kallus-Zhou2018, ding2022addressing, Wu-etal2022-framework}, which will impede valid inference about the target quantity such as the efficacy of a newly developed therapy.  This article aims to identify the drug effect on long-term outcomes in RCT by combining an RCT dataset and an observational dataset.   

In this paper, we mainly make the following three contributions.
	 First, under a mean exchangeability assumption, we elaborate on the identifiability of the treatment effect on long-term outcomes in RCT by combining an RCT without observation of long-term outcomes and an observational study in which the long-term outcome is observed but unmeasured confounding may exist. We
  % \bcol{compare our identifiability assumptions with those of existing methods and } 
  show that the identifiability assumptions adopted in this article are weaker than those of existing methods. %the previous literature. 
	 Second, we derive the efficient influence function and the semiparametric efficiency bound for the target parameter.  
	 Third, we propose a locally efficient doubly robust (DR) estimator and an inverse probability weighted (IPW) estimator, and show their large sample properties.   
	The proposed DR estimator is   
	  locally efficient in the sense that it attains the semiparametric efficiency bound if all the working models are correctly specified.   
However, the proposed DR estimator relies on the estimations of multiple complex nuisance parameters contained in the efficient influence function, its efficiency may degrade significantly if some of the nuisance parametric models are misspecified. This is not a problem unique to our method, existing approaches have similar problems~\citep[see][]{athey2016estimating, athey2020combining, kallus2020role, chen2021semiparametric}.  
To ease this problem, we further propose a simpler IPW estimator that relies on much fewer nuisance parameters and shows that using an estimated propensity score will lead to better performance even if we know the true propensity score, which is common for RCTs  \citep{robins1994estimation,hirano2003efficient}. 
    In addition, we show that the IPW estimator with estimated propensity scores achieves the semiparametric efficient bound in the case of discrete covariates and surrogates with finite support.   
  Both the proposed doubly robust and IPW estimators are shown to be consistent and asymptotically normally distributed, and the associated variance can be estimated. Extensive simulations are conducted to evaluate the performance of the proposed estimators.

The idea of leveraging external data to help identify and improve efficiency has gained much attention in the field of causal inference \citep{bareinboim2016causal,hunermund2019causal,kallus2018removing,yang2020elastic,yang2020improved,yang2020combining,li2021improving}. % Wu-etal2022-framework
Our method is closely related to the recently proposed methods of studying the long-term treatment effect.   
\citet{athey2016estimating} considered identifying the long-term causal effect in RCT data in a setting where the long-term outcome is not observed in RCT data and the treatment variable is missing in observational data.  
They briefly discussed estimation methods for the average treatment effect in RCT data, without showing the large sample properties of the proposed estimator. 
% \bcol{In addition, they .}
Different from their setting, we assume that treatment variables are observed in the observational study and allow treatment to have a direct effect on outcomes. We identify the same parameter under weaker assumptions than those in \citet{athey2016estimating}, derives the semiparametric efficiency bound, proposes two new estimators, and shows their asymptotic properties.  In addition, \citet{kallus2020role} considered the efficiency gain of estimating the causal effect on a long-term outcome by using an observed short-term surrogate when the long-term outcome is missing at random. However, the missing at random assumption is less plausible in the combined RCT and observational data.
% since it is inevitable to suffer from the problem of unmeasured confounders in observational data.

Several works have considered a causal parameter similar to ours, which is the long-term causal effects in observational data \citep{athey2020combining,ghassami2022combining,imbens2022long,chen2021semiparametric}, arguing that this quantity may have better generalizability.  But in practice, there are occasions when RCT is a representative sample of the target population, for example, \citet{li2021improving,athey2020combining} consider the average treatment effect in the RCT data for a new drug or a policy. 
Besides, real-world RCTs such as ``pragmatic randomized clinical trial" can contain samples reflective of real-world population \citet{gamerman2019pragmatic} and become more popular in recent years.  Furthermore, in empirical medical data analysis, the analysis results based on RCT data are more credible and are more easily accepted by regulators such as FDA.  Therefore, we choose the long-term causal effects in the RCT as the target parameter.
%  \citet{athey2020combining}  proposed latent unconfoundedness assumption to identify it,      
% \citet{ghassami2022combining} proposed two methods to obtain identifiability, which rely on an equi-confounding bias assumption or the existence of an extra proxy variable respectively, and 
% \citet{imbens2022long} considered using multiple surrogates that have a sequential structure to identify it.  
% Unlike our methodology, all these methods maintain the external validity assumption which states that conditional on covariates, the distribution of the potential surrogates and outcomes are the same in RCT data and observational data. Besides, these methods mainly focus on the average treatment effect in observational data and did not give the semiparametric efficiency bound for the target parameter,  
% except for \citet{chen2021semiparametric}, which complemented the semiparametric efficiency theory under the same setting as  
% \citet{athey2020combining}. Moreover, \citet{chen2021semiparametric} showed that the long-term causal effect in RCT can also be identified under the assumptions in \citet{athey2020combining}. In Section 2.3, we show that  identifiability assumptions are stronger than ours. 

The rest of the article is organized as follows. 
In Section 2, we describe the setting of the problem interested and give the identifiability assumptions,  and compare them with the existing approaches.  
Section 3 shows the semiparametric efficiency bound for the target parameter. Section 4 proposes two new estimators and presents their large sample properties. 
In Section 5, extensive simulations are performed to evaluate the finite sample behaviors of the proposed methods. Section 6 illustrates our approaches with an empirical example. A brief discussion is concluded in Section 7.

% ===========================================================================

\section{Causal Parameter and Identifiability}
\label{sec:identification}

\subsection{Study design and causal parameter} 

When combining datasets from different sources, sampling mechanisms of the multiple datasets are crucial for statistical inference. There are mainly two ways to view the study design of the RCT data and observational data: nested design and non-nested design \citep{colnet2020causal,dahabreh2021study}. In this paper, we adopt the non-nested design where the sampling mechanisms of the RCT are independent of the observational data.  Suppose that there exists an underlying population for the patients, and two subpopulations with different distributions. The RCT data and observational data are simple random samples from two corresponding subpopulations, where the sampling probabilities for the two subpopulations are unknown. With the observed data, the distributions of the two subpopulations can be identified, while the underlying population is not because of the unknown sampling probabilities. And the overall population for the observed data consists of samples from two subpopulations. A more detailed discussion of the related study design can be found in \citet{li2021improving}.
% \col{When combining datasets from different sources, sampling mechanisms of the multiple datasets are crucial for the statistical inference. there are mainly two ways to view the study design of the RCT data and observational data: nested design and non-nested design \citep{colnet2020causal,dahabreh2021study}. In this paper we adopt the non-nested design, which means that the sampling mechanisms of the RCT and observational data are assumed to be independent. To explicitly describe the observed data structure, we consider the following setup. Suppose that there exists an underlying populaition for the patients. The RCT data and observational data are simple random samples from two corresponding subpopulations, where the sampling probabilities for the two subpopulations are unknown. With the observed data, the distributions of the two subpopulations can be identified, while the underlying population is not because of the unknown sampling probabilities. And the overall population for the observed data consists of samples from two subpopulations. A more detailed discussion of the related study design can be found in \citet{li2021improving}. }

Now we introduce the observed data structure in our problem. 
Let $T$ denote the indicator for binary treatment, with $T=1$ or 0 the treated or control group, $X$ denotes the observed pre-treatment covariates,   $Y$ denotes the long-term outcome of interest, and $S$ denote the short-term surrogates (e.g., intermediate outcomes) that are highly informative about the outcome $Y$ and measured after the treatment $T$. Under the potential outcome framework \citep{Rubin1974, Neyman-1990},  let $\{S(1), Y(1)\}$ and $\{S(0),Y(0)\}$ be the potential outcomes  with and without treatment respectively.   
The observed surrogate $S$ and outcome $Y$ are the potential outcomes corresponding to the treatment received by the consistency assumption, i.e. $S = S(T)$ and $Y=Y(T)$. Suppose that we have available two data sources: an RCT dataset $\{ (T_{i}, X_{i}, S_{i}): i=1, ..., n_{1} \}$ consists of independent and identically distributed (i.i.d.) sample of $n_{1}$ observations, and an i.i.d. observational dataset  $\{ (T_{i}, X_{i}, S_{i}, Y_{i}) : \bcol{i=n_1+1, ..., n_1+n_{0}}\}$ contains $n_{0}$ observations. Therefore, the observed data has sample size $n = n_0+ n_1$.
 Denote $G_{i} \in \{0, 1\}$ as the indicator of the data sources, where $G_{i}=1$ represents that unit $i$ belongs to  RCT data and $G_{i}= 0$ represents that unit $i$ belongs to observational data. The limit of $n_1/n$ as $n\rightarrow \infty$ tends to a positive constant $q = \pr(G=1)$, which represents the proportion of the RCT data in the observed data population.
The parameter of interest is the average treatment effect in the RCT defined by $\tau = E\{Y(1) - Y(0) | G = 1\}$.

\subsection{Assumptions and identifiability} 
For identification, Assumptions 1 and 2 are imposed throughout. 

% \medskip 
\noindent
\emph{Assumption 1} (Internal validity of RCT data) % \label{Internal validity}
For $t = 0$ or 1,  
\[
T \ind (Y(t), S(t)) \mid X, G=1. 
\]

%\medskip 
\noindent
\emph{Assumption 2} (Strict overlap) % \label{overlap}
 There exists a constant $0 < \varepsilon < 1/2$, such that \par  
(i) $\varepsilon \leq e(X) := \pr(T=1\mid X,G=1) \leq 1 - \varepsilon$, \par
(ii) $\varepsilon \leq \pr(T=1\mid X,S,G=0) \leq 1 - \varepsilon$, \par
(iii) $\varepsilon \leq \pr(G=0 \mid X=x,S=s)$ for all $(x,s)$ satisfying  $\pr(X=x,S=s\mid G=1) > 0$.

Assumption 1 guarantees that the treatment assignment in RCT is unconfounded and is satisfied in most cases with a carefully designed experiment.  
\cite{kallus2020role} uses the assumption $T \ind (Y(t), S(t))\mid X$, where unconfoundedness holds in the combined data, rather than in RCT data. However, this assumption is less plausible, as the existence of unmeasured confounders is an unavoidable problem for observational data \citep{kallus2018removing}.  
Assumptions 2(i)-(ii) are common in causal inference literature \citep{Rosenbaum-Rubin1983, Tsiatis-2006, Imbens-Rubin2015, Hernan-Robins2020}, which ensure that each unit has the chance to be assigned to each treatment option.  
Assumption 2(iii) means that each unit in RCT has a positive probability of belonging to % the RCT group and
the observational data group. 
This implicitly restricts the support of covariates and surrogates in RCT data should be included in those of observational data, which is necessary to leverage observational data to help identify $\tau$.  
Besides, Assumption 2(iii) is reasonable in empirical studies because the inclusion rule exerted in RCT will prevent part of the patients from entering the experiment,  leading to a smaller support set of $X$ and $S$.    
The causal estimand $\tau$ is not identified under Assumptions 1 and 2,  we further invoke the following mean exchangeability assumption. 

% \medskip 
\noindent
\emph{Assumption 3} (Mean exchangeability)  % \label{Exchangability} 
For $t = 0$ or 1, 
\[ 
E(Y(t) \mid X, S(t), T=t, G=1) = E(Y(t) \mid X, S(t), T=t, G=0).
\] 

%  assumption
By consistency, Assumption 3 can be written as $E(Y|X, S, T, G=1)=E(Y| X, S, T, G=0)$, an equation only consists of observed data, which is the key assumption that enables us to transfer the conditional mean of $Y$ in observational data to RCT data. 
The mean exchangeability is a weaker version of $G\ind Y(t) \mid X,S(t),T=t$  brought up by \citet{kallus2020role} and similar assumptions are invoked across various data fusion literature \citep{li2021improving, wu2021integrative, miao2022invited}.    
 Importantly,  Assumption 3 allows for the existence of unmeasured confounders between the treatment $T$ and the surrogates $S$, as the unmeasured confounders have no direct effect on $Y$, leading to the same conditional expectation of $Y$  between the two datasets. 
Figures 1-2 give typical causal graphs when Assumption 3 holds. The following Proposition \ref{identification} gives the identification result for $\tau$. 

\begin{figure}  \label{DAG}
\centering
\begin{minipage}{0.4\textwidth}
\centering
\begin{tikzpicture}
\node (X) {$X$};
\node (S) [right =of X] {$S$};
\node (Y) [below=of S] {$Y$};
\node (T) [left=of Y] {$T$};
\node (U) [left=of T] {$U$};
\path (X)  edge (T);
\path (X)  edge (S);
\path (X)  edge (Y);
\path (T)  edge (S);
\path (T)  edge (Y);
\path (S) edge (Y);
% \path (Y) edge (S);
% \path (U)  edge (T);
\path (U)  edge (S);
% \path (D1) edge  [left = 45] node {$\beta_{1}$} (Y);
% \path (D2) edge node{$\beta_{2}$} (Y);
% \path (X2) edge [bend left = 45] node[below = 0.15 cm] {} (Y);
\end{tikzpicture}
\vspace{-0.5cm}
\caption{RCT data}     % caption
\end{minipage}
\begin{minipage}{0.4\textwidth}
\centering
\begin {tikzpicture}
\node (X) {$X$};
\node (S) [right =of X] {$S$};
\node (Y) [below=of S] {$Y$};
\node (T) [left=of Y] {$T$};
\node (U) [left=of T] {$U$};
\path (X)  edge (S);
\path (X)  edge (Y);
\path (X)  edge (T);
\path (T)  edge (S);
\path (T)  edge (Y);
\path (S)  edge (Y);
% \path (Y)  edge (S);
\path (U)  edge (T);
\path (U)  edge (S); 
% \path (U)  edge [bend right = 45] (Y);
% \path (D1) edge  [left = 45] node {$\beta_{1}$} (Y);
% \path (D2) edge node{$\beta_{2}$} (Y);
% \path (X2) edge [bend left = 45] node[below = 0.15 cm] {} (Y);
\end{tikzpicture}\vspace{-0.5cm}
 \caption{Observational data}     % caption
\end{minipage} 
\begin{flushleft}
Note: $U$ denotes the unmeasured confounder. 
\end{flushleft}
\end{figure}

% Let $h(x,s,t) = E(Y| X=x, S=s, T=t, G=1)$, which can be identified by $E(Y|X=x, S=s, T=t, G=0)$.    
\begin{proposition}\label{identification} 
Under Assumptions 1, 2 and 3,  $\tau$ is identified.
\end{proposition} 
% \begin{proof} It follows from the following equations 
% \begin{align}
% \tau &= E\left\{\frac{YT}{e(X)} - \frac{Y(1-T)}{1 - e(X)} \mid G=1\right\} \notag \\
% &= E\left\{\frac{h(X,S,T)T}{e(X)} - \frac{h(X,S,T)(1 - T)}{1 - e(X)}\mid G = 1\right\}. 
% \end{align}
% \end{proof}

\subsection{Comparison with the identifiability of existing methods} % assumptions 

The previous literature adopts stronger identifiability assumptions than Assumptions 1-3.  Concretely,  \cite{athey2020combining} and \cite{chen2021semiparametric} adopt the following Assumptions 4 and 5 to substitute Assumption 3.

% \medskip
 \noindent
\emph{Assumption 4} (Conditional external validity)  %\label{Conditional external validity}
\[ 
G_{i} \ind\left(Y_{i}(0), Y_{i}(1), S_i(0), S_i(1)\right) \mid X_{i}
\]

% \medskip
 \noindent
\emph{Assumption 5} (Latent unconfoundedness) % \label{Latent unconfoundedness} 
For $t=0$ or 1, 
\[
T_{i} \ind Y_{i}(t) \mid X_{i}, S_i(t), G_{i}=0
\]

Under Assumptions 1, 2, 4, and 5, \citet{athey2020combining} and \citet{chen2021semiparametric} obtain the identifiability of the average treatment effect in observational data, i.e.,  $E[Y(1) - Y(0) | G=0 ]$, and the authors assert that these assumptions are also applicable to identify $\tau$.   
However, Assumptions 4--5 may be too strong for empirical applications when the focus is the average treatment effect in RCT data.  Assumption 4 states that conditioning on $X$, the distributions of potential outcomes are the same between RCT data and observational data, which implicitly assumes that the distributions of the unmeasured confounders affecting $T$ and $S$   conditional on observed covariates are the same between RCT data and observational data. 
Compared to Assumptions 4--5,   Assumption 3 imposes weaker constraints on the data-generating distribution. In fact, we can show that Assumptions 4--5 are sufficient conditions for Assumption 3 under Assumption 1.

\begin{proposition}  Under Assumption 1, Assumption 3 is implied by  Assumptions 4 and 5.  
\end{proposition}  

Below we provide an example that satisfies Assumption 3 but violates Assumptions 4-5.  
\begin{example}\label{example1}  
Consider the following structural equation models. 
For RCT data:  
\[
\begin{aligned}
\pr(T=1) &= 1/2, S = \alpha_1 X + \alpha_2 U+ \tau_S T + \varepsilon_{S}\\ Y&= \beta_1 X +\beta_2 S+ \tau T + \varepsilon_{Y} \\ 
\end{aligned}
\] 
For observational data:
\[
\begin{aligned}
\pr(T=1 \mid X, U) &= \{ 1 + \exp( - \gamma_1 X - \gamma_2 U)\}^{-1}, S = \alpha_1 X + \alpha_2 U+ \tau_S T + \varepsilon_{S} \\ 
Y&= \beta_1 X + \beta_2 S+ \tau T + \varepsilon_{Y} \\ 
\end{aligned}
\]
where $U$ is an unmeasured variable in both RCT data and observational data,  $\varepsilon_{S}$ and $\varepsilon_{Y}$ be the error terms independent of all other variables.   
 If the distribution of  $U|G=1$ and $U|G=0$ are different, one can verify that the distribution of $S(t), Y(t), t=0,1$ are different in RCT data and observational data, thus Assumption 4 is violated. 
 \end{example}

 % =========================================================================
\section{Semiparametric Efficiency Bound}
\label{sec3}

Under the nonparametric model restricted by Assumptions 1--3, we calculate the semiparametric efficiency bound for $\tau$.   
 The following intermediate quantities will appear in the efficient influence function: 
   the selection propensity score $g_t(s,x) = \pr(G=1|S=s,X=x,T=t)$, which quantifies the probability of selection into RCT group for a given surrogate, baseline covariates and treatment;    
the treatment propensity score for RCT  $e(x) = \pr(T=1| X=x, G=1)$;   
the regression functions  
$\mu_t(s, x) = E(Y(t) | S(t)=s, X=x, G=1) = E(Y| S=s, X=x, T=t, G=1)$ and $\mu_t(x) = E(Y(t) | X=x, G=1) = E(Y| X=x, T=t, G=1) = E\{\mu_t(S,X)|X=x, T=t, G=1\}$ for $t = 0, 1$. With these nuisance parameters, Theorem \ref{efficiency bound}  presents the efficient influence function  for $\tau$. 
% and efficiency bound 

\begin{theorem}[efficiency bound] \label{efficiency bound}
Under Assumptions 1--3, the efficient influence function for $\tau$ is given as  
\begin{align*}
\phi &= \frac{G}{q}\left\{ \frac{T(\mu_1(S, X) - \mu_1(X))}{e(X)} - \frac{(1-T)(\mu_0(S, X) - \mu_0(X))}{1 - e(X)} + (\mu_1(X) - \mu_0(X) )- \tau \right\} \\ 
&+ \frac{1-G}{q}\biggl\{\frac{g_1(S,X) T\{Y - \mu_1(S,X) \}}{e(X)\{1 - g_1(S,X) \}} -  \frac{g_0(S,X)(1-T)\{Y- \mu_0(S,X)\}}{\{1-e(X)\}\{ 1 - g_0(S,X) \}} \biggl\}, 
\end{align*}
where $q = p(G=1)$. The semiparametric efficiency bound is $ E(\phi^2)$. In addition, \par
	(i) the efficiency bound remains the same no matter whether the propensity score $e(X)$ is known or not. \par
  (ii) $\phi$ is the unique influence function in the nonparametric model class that is only restricted by Assumptions 1--3. 
\end{theorem}

Theorem \ref{efficiency bound} shows that for any regular and asymptotic linear estimator, its asymptotic variance is no smaller than the efficiency bound $E(\phi^2)$.   
\citet{chen2021semiparametric} obtains the efficient influence function for average treatment effect for the long-term outcome in observational data under Assumptions 1, 2, 4, and 5. Here our focus is the average treatment effect in RCT data, and the efficient influence function is derived under weaker assumptions. 
\bcol{Theorem \ref{efficiency bound}(i) shows that the propensity score is ancillary for the estimation of $\tau$, that is, the knowledge of $e(x)$  does not decrease the efficiency bound of $\tau$.}
% As the parameter of interest is not a functional of $e(X)$, the efficiency bound is not influenced by the knowledge of the propensity score $e(X)$, as shown in Theorem \ref{efficiency bound}(i).   
Similar conclusions that knowing some nuisance parameters will not change the efficiency bound of the target parameter are made in \cite{hahn1998role} and \cite{chen2021semiparametric}.  
The uniqueness of the influence function in Theorem \ref{efficiency bound}(ii) means that any regular and asymptotic linear estimators for $\tau$ in the nonparametric model have the same influence function and thus the same asymptotic distribution.

 % ======================================================================
\section{Estimation}
\label{sec4}

\subsection{Efficient doubly robust  estimator} 

%The efficient influence function $\phi$ in 
Theorem~\ref{efficiency bound} motivates an estimator that can achieve the semiparametric efficiency bound. 
Concretely,  generalized linear models are specified for the nuisance parameters in $\phi$, including $\mu_t(s,x; \alpha_t)$ and $\mu_t(x;\beta_t)$ for $t=0$ or 1, $e(x;\gamma)$, and $g_t(s,x;\eta_t)$. 
Let $\hat \alpha_{t}$, $\hat \beta_{t}$, $\hat \gamma$, and  $\hat  \eta_t$  \bcol{denote} the maximum likelihood estimators of $\alpha_{t}$, $\beta_{t}$, $\gamma$, $  \eta_t$, respectively.  

 The estimation of $e(x;\gamma)$ is trivial. However,  
particular care is needed when estimating $g_t(s,x;\eta_t)$, $\mu_{t}(s, x; \alpha_{t})$ and $\mu_{t}(x; \beta_{t})$. First,  $g_t(s,x;\eta_t)$ should be estimated based on  
the combined samples of both RCT and observational data; Second, 
 we cannot estimate $\mu_{t}(s, x; \alpha_{t})$ directly due to the missingness of $Y$ in RCT data.  
 Owing to Assumption 3, $\mu_{t}(s, x; \hat \alpha_{t})$ can be obtained by regressing $Y$ on $(X, S)$ in  observational data with $T = t$, then we calculate their predicted values in RCT data ; Finally,  
    $\mu_{t}(x; \hat \beta_{t})$ can be derived by conducting a linear regression of $\mu_{t}(s, x; \hat \alpha_{t})$ on $X$ in the RCT sample.  
     It should be noted that we can't estimate $\mu_{t}(x; \beta_{t})$ by directly regressing $Y$ on $X$ with  observational data, since $E[Y|X, T=t, G=1] $ may not equal to $E[Y|X, T=t, G=0]$ under Assumptions 1--3.  
With these fitted nuisance parameters, the efficient doubly robust estimator is given as   
\[
\begin{aligned}
\hat{\tau}_{dr} &= \hat{E}\bigg[ \frac{G}{\hat{q}}\bigg\{ \frac{T(\mu_1(S, X;\hat{\alpha}_1) - \mu_1(X;\hat{\beta}_1))}{e(X; \hat{\gamma})} - \frac{(1-T)(\mu_0(S, X;\hat{\alpha}_0) - \mu_0(X;\hat{\beta}_0))}{1 - e(X;\hat{\gamma})} + \mu_1(X;\hat{\beta}_1) \\ 
 -& \mu_0(X;\hat{\beta}_0) \bigg\}  + \frac{1-G}{\hat{q}}\biggl\{\frac{g_1(S,X;\hat{\eta}_1) T\{Y - \mu_1(S,X) \}}{e(X;\hat{\gamma})\{1 - g_1(S,X;\hat{\eta}_1) \}} -  \frac{g_0(S,X;\hat{\eta}_0)(1-T)\{Y- \mu_0(S,X)\}}{\{1-e(X;\hat{\gamma})\}\{ 1 - g_0(S,X;\hat{\eta}_0) \}} \biggl\} \biggl ],
\end{aligned}
\]
where $\hat E(\cdot)$  denotes the sample average of all data throughout, $\hat q = n_{1} / (n_{1} + n_{0})$. 
The large sample properties of $\hat \tau_{dr}$ are presented in the following Theorem \ref{th2}.

\begin{theorem}\label{th2}
Under Assumptions 1--3 and regularity conditions described in theorems 2.6 and 3.4 of \citet{newey1994large}, the estimator $\hat{\tau}_{dr}$ is consistent and asymptotically
normal if either 

(i) the outcome model $\mu_t(S,X;\alpha_t)$ and $\mu_t(X;\beta_t)$
for $t = 0,1$ are correctly specified, or

 (ii) the outcome model $\mu_t(S,X;\alpha_t)$ and the propensity score model $e(X;\gamma)$ are correctly specified. 

In addition, $\hat \tau_{dr}$ is locally efficient, i.e., it attains the semiparametric efficiency bound $E(\phi^{2})$ when all the working models are correctly specified.
\end{theorem} 

Theorem \ref{th2} indicates that the consistency of $\hat{\tau}_{dr}$ relies on the correct specifications of $\mu_t(S,X)$ for $t=0,1$, which may not be guaranteed in real data analysis and increases the risk of obtaining biased conclusions. However, the consistency of $\hat{\tau}_{dr}$ does not rely on the correct specification of the selection propensity score $g_t(s,x)$, although the asymptotic variance does. Besides,  the doubly robust estimator involves many nuisance parameters and some of them are intertwined. For example, the definitions of  $\mu_t(S, X)$ and $\mu_t(X)$ imply that $\mu_{t}(X) = E[ \mu_{t}(S, X) | X, G=1 ]$. When a logistic model is specified for $\mu_t(S, X)$, a logistic model for $\mu_t(X)$ can hardly be correctly specified. We found that this is not a problem unique to our method, existing approaches have similar problems~\citep[see][]{athey2016estimating, athey2020combining, kallus2020role, chen2021semiparametric}.
    
When all the parametric models for the nuisance parameters are correctly specified, the asymptotic variance of $\hat{\tau}_{dr}$ can be naturally estimated by $\hat{E}(\hat \phi^2)$, where  $\hat \phi$ is the plug-in estimator of $\phi$.  
Besides, we can use the bootstrap method to get the asymptotic variance estimation if we cannot ensure the correctness of all model specifications. Concretely, for each bootstrap, we randomly sample $n_1$ and $n_0$ samples from RCT and observational data with replacement, respectively. Repeat $B$ times to get $B$ point estimates. Then the sample variance of the $B$ point estimates is the estimate of the asymptotic variance of $\hat \tau_{dr}$.  
 In Section 5, our simulation compares these two methods of calculating the asymptotic variance.

\subsection{Inverse probability weighted estimator}  
 
 The doubly robust estimator has some worrying features. As discussed in Section 4.1, its efficiency relies on the correctness of multiple cumbersome model specifications for the nuisance parameters.   
 When some models are misspecified, the efficiency may degrade  and the estimator may have a bias.      % significantly
As a complement, the inverse probability weighted (IPW) estimator that can consistently estimate $\tau$ by merely imposing a model specification for $h(X, S, T) = E[Y|X, S, T, G=1 ]$, which is given by  
\begin{equation} \label{eq-ipw1}
\hat \tau_{ipw} = \frac{1}{n_1 }	\sum_{i=1}^{n_1} \biggl \{   \frac{T_{i}  \cdot  h(X_i, S_i, T_i; \hat \kappa)  }{  e(X_{i})}   -  \frac{ (1 - T_{i}) \cdot  h(X_i, S_i, T_i; \hat \kappa) }{1-  e(X_{i})}  \biggr \},
\end{equation}
 where $h(X, S, T; \kappa)$ is assumed to be a generalized linear model and $\hat  \kappa$ is the maximum likelihood estimator of $ \kappa$ based on observational data. Clearly, the IPW estimator has a much simpler form than the doubly robust estimator and thus is more tractable.

 Generally,  there are two obstacles to applying the IPW estimator in statistical analysis: imprecision and instability when some propensity score values are close to 0 or 1 \citep{Tan2007,  Tan2010, Molenberghs-etal2015, Wu-etal-2021, Wu-Han2021}. 
   Since the propensity score in RCT data is usually known and bounded away from 0 or 1, the problem of instability does not exist in our setting. 
      To improve the efficiency of the IPW estimator, we propose using
 the estimated propensity score, instead of the true propensity score, to construct the IPW estimator. Specifically,  we use logistic regression to estimate it, i.e.,  assuming 
$e(X_i) = e(X_i; \gamma) =  \exp( X_{i}^{T} \gamma ) / \{ 1 + \exp(X_{i}^{T} \gamma ) \}$. Let  $\hat \gamma$ be the maximum likelihood estimator of $\gamma$, and define  
\begin{equation} \label{eq-ipw2}
\tilde \tau_{ipw} =   \frac{1}{n_1 }	\sum_{i=1}^{n_1} \biggl \{   \frac{T_{i}  \cdot  h(X_i, S_i, T_i; \hat  \kappa)  }{  e(X_{i}; \hat \gamma)}   -  \frac{ (1 - T_{i}) \cdot  h(X_i, S_i, T_i; \hat  \kappa) }{1-  e(X_{i}; \hat \gamma)}  \biggr \}. 
\end{equation}

%  \begin{assumption} \label{assump6} $n_1 / n_0 \rightarrow \rho$ as both $n_{1}$ and $n_{0}$ go to infinity, where $\rho$ is a constant.    
% \end{assumption} 

%  The constant $\rho$ balances the efficiency contributions of RCT data and observational data to the IPW estimator.  
Next, we establish the asymptotic properties of $\hat \tau_{ipw}$ and $\tilde \tau_{ipw}$.
For ease of exposition hereafter, we let
 $e_{i} = e(X_{i})$,  $\hat e_{i} = e(X_i; \hat \gamma)$,    
 $\tilde X_{i} = (X_i^T, S_i^T, T_i)^{T}$,  $h_{i} = h(X_i, S_i, T_i; \kappa)$,  $\hat h_{i} = h(X_i, S_i, T_i; \hat \kappa)$, and $h_i'(\kappa) = \partial h(X_i, S_i, T_i; \kappa) / \partial \kappa$. Denote the true values of $\kappa$ and $\gamma$ by $\kappa^*$ and $\gamma^*$.

\begin{theorem} \label{thm3}  Under Assumptions 1--3, and denote $\rho = p(G=1)/\{1 - p(G=1)\}$, which is the limit of $n_1 / n_0$, we have

(i) if the propensity scores $e_{i}$'s in RCT data are known, then 
      \begin{equation} \label{asy1}  \sqrt{n_1} ( \hat \tau_{ipw} - \tau )  \xrightarrow[]{d} N \Big (0, 
      V_1 + \rho  B_1^T I^{-1}(\kappa^*) B_1  \Big  ),     \end{equation}
    where $V_1 = \text{var}\{ \frac{(T_i - e_{i} ) h_i}{ e_{i} (1 - e_{i} ) } | G_i = 1 \}$, 
    $B_1 = E[ \frac{ (T_{i}-e_{i})   }{  e_{i} (1 - e_{i}) } \cdot  h_i'(\kappa^*)  | G_{i} = 1] $ and  $I(\kappa^*)$ is the Fisher information matrix of $\kappa$ at $\kappa^*$ in observational data.  
    
(ii)  % \bcol{if the logistic regression model for propensity score is correctly specified}, then 
if we estimate the propensity scores $e_{i}$'s in RCT data with a correctly specified logistic regression model, then 
\begin{equation}\label{asy2}  \sqrt{n_1}  ( \tilde \tau_{ipw} - \tau )  \xrightarrow[]{d} N \Big (0, 
       (V_1- V_2)  +  \rho B_1^T I^{-1}(\kappa^*) B_1  \Big  ),     \end{equation}
    where $V_2 =  B_2^{T} I^{-1}(\gamma^*) B_2,$ 
with $B_2 = E[ \frac{ T_i h_i (1-e_i)X_i }{e_i} | G_i = 1 ] +  E[ \frac{ (1-T_i) h_i e_i X_i }{ 1 - e_i } | G_i = 1  ]$,  
    $I(\gamma^*)= E[ e_{i}(1 - e_{i})  X_{i} X_{i}^{T} | G_i = 1 ]$ is the Fisher information matrix of $\gamma$ at $\gamma^*$. 
\end{theorem}

Theorem \ref{thm3}(i) shows the asymptotic variance of $\sqrt{n_{1}} \hat \tau_{ipw}$ consists of $V_1$  and $\rho B_1^T I^{-1}(\kappa^*) B_1$, where the former is the variance of IPW estimator when $h(X, S, T)$ is known  by noting that $V_{1} = \text{var}\{ T_{i} h_{i} / e(X_{i})   -  (1 - T_{i}) h_{i}/(1-  e(X_{i})) | G_i = 1 \}$, which can be seen as the systematic variance; the latter is induced by the estimation of $h(X_i, S_i, T_i; \kappa)$.  Compared with $\hat \tau_{ipw}$,  the asymptotic variance of $\tilde \tau_{ipw}$ in Theorem \ref{thm3}(ii)  minus an extra positive term $V_{2}$ resulted from the estimation of propensity scores, which reveals that using estimated propensity scores reduces the asymptotic variance and thus lead to a more accurate estimator. This phenomenon has been noticed in previous literature, such as \citep{Joeef-Rosenbaum1999, hirano2003efficient, Wu-etal-2021}, and we will verify it in the simulation study of Section \ref{sec5}.    

The results given in Theorem \ref{thm3} are valid for any generalized linear model $h(X_i, S_i, T_i; \kappa)$, and thus it is applicable to various data types of $Y$. For convenience,  we present the specific form of $B_{1}$ and 
   $I(\kappa^*)$ for binary and continuous outcomes, the two most common scenarios in real data analysis.  (1) For binary $Y$ and assume $h(X_i, S_i, T_i; \kappa)$ is a logistic model, then 
$B_1 = E[ \frac{ (T_{i}-e_{i})   }{  e_{i} (1 - e_{i})  } \cdot h_{i} (1 - h_{i}) \tilde X_{i} | G_{i} = 1] $,  $I(\kappa^*) = E[ h_{i}(1 - h_{i}) \tilde X_{i} \tilde X_{i}^{T} | G_i = 0 ]$. (2) For continuous outcome and assume $h(X_i, S_i, T_i; \kappa)$ is a linear model with variance $\sigma^{2}$, then 
     $B_1 = E[ \frac{ (T_{i}-e_{i})   }{ e_{i} (1 - e_{i}) } \cdot  \tilde X_{i} | G_{i} = 1] $,  $I(\kappa^*) = E[ \tilde X_i \tilde X_i ^T \sigma^{-2} |G_i = 0]$.

Furthermore, with respect to the efficiency between the IPW and the efficient doubly robust estimator, we have the following corollary.
\begin{corollary} \label{coro}
When $(X,S)$ are discrete with finite support, \bcol{and the nuisance parameters in $\hat{\tau}_{dr}$ and  $\tilde{\tau}_{ipw}$ are nonparametrically estimated of order $\sqrt{n}$}, as the efficient influence function $\phi$ is the unique influence function for $\tau$, we have $\hat{\tau}_{dr}$ and  $\tilde{\tau}_{ipw}$ are first-order equivalent, that is, they have the same asymptotic distribution.
\end{corollary}

Corollary \ref{coro} shows that the IPW estimator using estimated propensity scores achieves the semiparametric efficient bound in the case of discrete $X$ and $S$ with finite support.  
The intuition is that when $X$ and $S$ are discrete with finite support, the selection model $g_t(S,X)$, the propensity score model $e(X)$, and the regression model $\mu_t(S,X), \mu_t(X)$ only contain finite dimensional parameters, which can be nonparametrically estimated \bcol{at a convergence rate of order $1/\sqrt{n}$}. Therefore the IPW method with estimated propensity score and the doubly robust estimator is regular and asymptotically normal for the nonparametric model constrained only by Assumptions 1--3. Their corresponding influence function must be the only element $\phi$ \bcol{by Theorem ~\ref{efficiency bound}(ii)}, so the two estimators have the same asymptotic distribution. 

The asymptotic variances of both $\hat \tau_{ipw}$ and $\tilde \tau_{ipw}$ can be obtained by the plug-in method, that is, substitute $e(X_{i})$ and $h_{i}$ with its estimates  $\hat e(X_{i})$ and $\hat h_{i}$ in the associated asymptotic variance formulas, and the population expectation is replaced by the empirical average.   
 In Section \ref{sec5}, the \bcol{simulation} study shows that the estimated asymptotic variances based on the plug-in method perform well across extensive simulation scenarios.

% ===============================================================

\section{Simulation}
\label{sec5}

We conduct extensive simulation studies to assess the finite sample performance of the proposed methods and compare them with the competing approach of \cite{athey2016estimating}.  Two common data types of $Y$, binary and continuous, are considered in this simulation.  
\texttt{R} codes are provided in \url{https://github.com/hwj0828/long-term-effect} to reproduce the simulation results.
% Supplementary Materials

Denote $U$ as the unmeasured variable. 
Throughout this simulation, for RCT data, unmeasured variable $U \sim N(0,1)$, and the treatment assignment  $\pr(T=1) = 1/2$.  The error terms $\varepsilon_S$ and $\varepsilon_Y$ are independently identically distributed in $N(0, 1)$ for both the RCT and observational data. 
The sample size of RCT data is set as $n_{1} = 50, 100$ or 200, observational data is $n_{0} = 500$.    Let $\text{expit}(x) = \exp(x)/\{1 + \exp(x) \}$ be the logistic function.

\medskip \noindent 
{\bf Continuous outcome.}    We first consider the following four cases for continuous $Y$.   

{\bf Case (1)}.  For RCT data ,  $S = U + 2(X_{1}+X_{2}) + T + \varepsilon_S$,  $ Y  =  T + 3(X_{1} + X_{2}) + S + \varepsilon_Y$, $X = (X_{1}, X_{2})^{T} \sim N(0,I_2)$.   
       For observational data,  $X \sim N(1, 4 I_2), U \sim N(0,1)$,  $\pr(T=1| X, U) = \text{expit}\{ U + X_{1} + X_{2} \}$, $S$ and $Y$ are generated the same as in RCT data . 
     
    {\bf Case (2)}. For RCT data , $S = U^2 + 2(X_{1}^2 + X_{2}^{2})  + T + \varepsilon_S,$  $Y  =  T + 3(X_{1} + X_{2}) + S + \varepsilon_Y$, $X = (X_{1}, X_{2})^{T} \sim N(0, I_2)$.  
      For observational data,  $X\sim N(1, 4 I_2), U \sim N(0,1)$,   $\pr(T=1| X, U) = \text{expit}\{ U + X_{1} + X_{2} U\}$, $S$ and $Y$ are generated the same as in RCT data . 
     
{\bf Case (3)}.  The data generation mechanism is the same as in case (1), except for setting $U \sim N(1, 4)$ in observational data. 
     
    {\bf Case (4)}.  The data generation mechanism is the same as in case (2), except for setting  $U \sim N(1, 4)$ in observational data. 
    
\vskip 0.2cm     
 
The unmeasured confounder $U$ influences both $T$ and $S$ in observational data for all cases (1)-(4).    
 The distribution of $U$ in RCT data is the same as that in observational data for cases (1) and (2) and is different from that in observational data for cases (3) and (4).   
 As discussed in Section 2, Assumptions 1-3 hold for all cases (1)-(4), while Assumption 4 holds only for cases  (1)-(2) and is violated for cases (3)-(4).  
In addition, to mimic the real-world data, we set the distributions of covariates between the RCT and observational data to be different for all cases (1)-(4), and the surrogates  $S$ may be a linear function of $(T, X, U)$ (cases (1) and (3)), or a  non-linear function of $(T, X, U)$ (cases (2) and (4)).  

 Each simulation study is based on 1000 
replicates. In the following tables,  Bias and SD are the Monte Carlo bias and    
standard deviation over the 1000 simulations of the points estimates. 
 ESE and CP95 are the averages of estimated asymptotic standard error and coverage proportions of the 95\% confidence intervals based on the plug-in method, respectively. ESE.b and CP95.b have the same meaning as ESE and CP95 but are derived from 200 bootstraps.  
 The true value of $\tau$ is obtained by generating RCT data with a sample size of 100000.

 \begin{table}
\centering
\caption{Comparison of various estimators for cases (1)-(4),  continuous outcome. The distributions of $U$ between the RCT and observational data are the same for cases (1)-(2) and different for cases (3)-(4).} 
\resizebox{\linewidth}{!}{\begin{tabular}{c rcccc  rcccc rcccc } 
\hline  	\hline
	 &    \multicolumn{5}{c}{$n_1 =50$}    & \multicolumn{5}{c}{$n_1 =100$} &  \multicolumn{5}{c}{$n_1 =200$}  \\
Case    & Bias (SD) & ESE & CP95 & ESE.b & CP95.b  & Bias (SD) & ESE & CP95 & ESE.b & CP95.b    & Bias (SD) & ESE & CP95 & ESE.b & CP95.b  \\  \hline  
			 &   \multicolumn{15}{c}{ 	\multirow{2}{*}{IPW Estimator ($\hat \tau_{ipw}$), with True Propensity Score} }    \\
	 &    \multicolumn{15}{c}{ }    \\
  (1)  &  8.1 (210.5) & 206.2 & 93.3 & 202.3 & 92.7 & 8.7 (144.8) & 146.6 & 94.9 & 145.4 & 94.0 & -2.1 (105.2) & 103.5 & 93.7 & 102.5 & 92.8  \\
  (2) & 1.7 (240.0) & 239.3 & 94.7 & 235.9 & 94.2 & -7.4 (176.4) & 171.3 & 93.9 & 169.3 & 93.1 & 2.1 (119.5) & 121.2 & 95.8 & 121.0 & 95.8  \\
  (3) &   13.1 (206.6) & 206.6 & 93.9 & 202.9 & 93.6 & 1.1 (147.5) & 145.2 & 95.3 & 144.1 & 95.0 & 9.8 (100.2) & 103.4 & 95.7 & 102.2 & 95.2  \\
  (4) &   7.0 (250.0) & 239.5 & 93.9 & 236.3 & 93.3 & 13.6 (168.2) & 171.4 & 95.9 & 169.8 & 95.1 & 8.2 (120.1) & 121.6 & 94.9 & 121.0 & 94.5  \\
	\hline  
	 &    \multicolumn{15}{c}{ 	\multirow{2}{*}{IPW Estimator ($\tilde \tau_{ipw}$), with Estimated Propensity Score} }    \\
	 &    \multicolumn{15}{c}{ }    \\
	(1) & 5.0 (51.6) & 55.6 & 96.4 & 92.1 & 98.6 & 2.9 (32.4) & 33.9 & 96.1 & 39.4 & 97.1 & 2.1 (23.9) & 24.4 & 95.0 & 25.4 & 95.1 \\
	(2) & 1.2 (138.8) & 125.9 & 91.4 & 163.7 & 95.4 & 0.7 (94.1) & 89.5 & 94.1 & 97.4 & 95.4 & 2.4 (66.2) & 63.5 & 94.2 & 65.4 & 94.7 \\
	(3) & 8.0 (52.1) & 55.6 & 96.4 & 95.8 & 99.4 & 7.0 (32.3) & 34.4 & 95.3 & 40.1 & 97.2 & 8.4 (23.8) & 24.8 & 94.3 & 25.8 & 93.6 \\ 
	(4) &  3.2 (150.0) & 128.0 & 92.3 & 174.9 & 95.8 & 7.9 (95.2) & 89.5 & 94.3 & 97.4 & 95.7 & 7.6 (66.0) & 63.6 & 93.9 & 65.3 & 94.0  \\
	\hline			
	 &    \multicolumn{15}{c}{ 	\multirow{2}{*}{Doubly Robust Estimator ($\hat \tau_{dr}$)} }    \\
	 &    \multicolumn{15}{c}{ }    \\
	(1) & -1.5 (41.9) & 48.7 & 96.8 & 39.4 & 91.8 & 0.3 (30.9) & 35.9 & 97.4 & 30.3 & 94.3 & 0.8 (23.8) & 28.0 & 97.4 & 24.2 & 94.9  \\
	(2) & -4.7 (116.7) & 118.1 & 94.8 & 104.0 & 91.6 & -1.4 (86.3) & 87.5 & 95.9 & 80.9 & 92.9 & 0.9 (63.9) & 63.8 & 95.3 & 60.7 & 93.3  \\
	(3) & 2.3 (41.2) & 49.0 & 98.2 & 39.5 & 93.2 & 4.9 (30.7) & 37.1 & 98.6 & 31.1 & 95.4 & 6.9 (25.3) & 30.7 & 98.4 & 25.5 & 93.6  \\
	(4) & -0.6 (123.3) & 119.3 & 95.1 & 104.9 & 91.4 & 4.1 (88.1) & 88.8 & 95.6 & 82.2 & 93.2 & 5.9 (65.8) & 65.4 & 95.3 & 61.9 & 92.5  \\
	\hline
 		 &   \multicolumn{15}{c}{ 	\multirow{2}{*}{\cite{athey2016estimating}'s Method} }    \\
	 &    \multicolumn{15}{c}{ }    \\
 	(1) & -83.2 (214.4) & 212.1 & 93.6 & 206.3 & 92.0 & -83.1 (147.3) & 149.9 & 91.9 & 148.2 & 92.0 & -93.9 (108.0) & 106.4 & 86.0 & 106.7 & 85.4 \\
	(2) & -101.2 (242.7) & 243.5 & 92.9 & 238.6 & 91.9 & -112.1 (178.0) & 173.5 & 88.6 & 171.2 & 87.1 & -108.9 (121.1) & 122.5 & 84.1 & 122.2 & 83.8  \\
	(3) & -80.7 (210.1) & 211.5 & 92.1 & 206.2 & 92.0 & -94.3 (149.6) & 148.2 & 89.3 & 147.3 & 88.2 & -90.4 (124.9) & 114.6 & 87.1 & 124.2 & 86.8 \\
	(4) & -99.5 (254.5) & 245.3 & 91.6 & 240.8 & 89.7 & -99.0 (170.8) & 175.0 & 91.8 & 173.2 & 91.0 & -115.4 (123.0) & 124.9 & 83.3 & 124.9 & 82.9   \\
	\hline		
	\hline
\end{tabular}}
\begin{flushleft}
{\footnotesize Note: All the values in this table have been magnified 100 times. Bias and SD are the Monte Carlo bias and   
standard deviation over the 1000 simulations of the points estimates. 
 ESE and CP95 are the averages of estimated asymptotic standard error and coverage proportions of the 95\% confidence intervals based on the plug-in method, respectively. ESE.b and CP95.b have the same meaning as ESE and CP95 but are derived from 200 bootstraps.}
\end{flushleft}
\label{tab-r1}
\end{table}

Table \ref{tab-r1} summarizes the numeric results of the proposed estimators (doubly robust estimator $\hat \tau_{dr}$, IPW estimators $\hat \tau_{ipw}$ and $\tilde \tau_{ipw}$) and the competing estimator in \cite{athey2016estimating} for cases (1)-(4). 
 For all the proposed estimators $\hat \tau_{dr}$, $\hat \tau_{ipw}$ and $\tilde \tau_{ipw}$,  
 the Bias is small, ESE is close to SD and CP95 is close to its nominal value of 0.95. This shows the validity of the asymptotic variance estimation using the plug-in method.  
As expected,  $\hat \tau_{dr}$ and $\tilde \tau_{ipw}$ have better performance than the $\hat \tau_{ipw}$ in terms of smaller Bias and SD. 
Remarkably, the results of the IPW estimator with estimated propensity score are similar to those of the doubly robust estimator.   
 In addition, the method of \cite{athey2016estimating}  has a significantly larger Bias than the other three approaches,  and its CP95 is less than 0.95. \bcol{A possible reason is that \cite{athey2016estimating}'s method does not allow $T$ to have a direct effect on $Y$, whereas in our setups we set $T$ to have a direct effect on $Y$.}

In the setup of simulation study, we assume that the treatment has a direct effect on the outcome. However,  , see Assumption 2 and the associated discussion in \cite{athey2016estimating}.

To verify the conclusion of Corollary \ref{coro}, we conduct two additional simulation scenarios (cases (5)-(6)).    
The data generating process of cases (5)-(6) are the same as cases (1)-(2), respectively, except that setting $X$ and $S$ as discrete variables. 
 For cases (5)-(6), the covariates $X$ consisting of two binary variables $(X_{1}, X_{2})$, $X_{1}$ and $X_{2}$ are independent and identically distributed from a Binomial distribution $B(1, 0.5)$ for both the RCT and observational data. The surrogate $S$ is generated through a logistic regression with $\pr(S = 1 | X, S, U, T) = \text{expit}\{U - 2(X_{1}+X_{2}) + T\}$ for case (5) and $\pr(S = 1 | X, S, U, T) = \text{expit}\{U^2 - 2(X_{1}^2 + X_{2}^{2})  + T\}$ for case (6).    
 Table \ref{tab-r2} shows the  simulation results for cases (5)-(6). As expected, the IPW estimator with estimated propensity scores and the doubly robust estimator have similar performance for discrete $X$ and $S$.   

\begin{table}
\centering
\caption{Comparison of various estimators for cases (5)-(6),  continuous outcome, discrete $X$ and $S$.}  
\resizebox{\linewidth}{!}{\begin{tabular}{c rcccc  rcccc rcccc } 
\hline  	\hline
	 &    \multicolumn{5}{c}{$n_1 =50$}    & \multicolumn{5}{c}{$n_1 =100$} &  \multicolumn{5}{c}{$n_1 =200$}  \\
Case    & Bias (SD) & ESE & CP95 & ESE.b & CP95.b  & Bias (SD) & ESE & CP95 & ESE.b & CP95.b    & Bias (SD) & ESE & CP95 & ESE.b & CP95.b  \\  \hline  
			 &   \multicolumn{15}{c}{ 	\multirow{2}{*}{IPW Estimator ($\hat \tau_{ipw}$), with True Propensity Score} }    \\
	 &    \multicolumn{15}{c}{ }    \\
(5) & -1.3 (122.1) & 120.4 & 94.2 & 118.9 & 93.2 & 1.2 (85.1) & 85.7 & 94.6 & 85.2 & 94.3 & -1.4 (58.5) & 61.0 & 96.0 & 60.6 & 95.0   \\
(6) &  8.2 (123.0) & 123.7 & 94.4 & 121.9 & 93.4 & 4.1 (88.6) & 87.8 & 93.7 & 86.8 & 92.5 & -0.9 (63.5) & 62.6 & 94.6 & 62.1 & 94.1  \\
	\hline
	 &    \multicolumn{15}{c}{ 	\multirow{2}{*}{IPW Estimator ($\tilde \tau_{ipw}$), with Estimated Propensity Score} }    \\
	 &    \multicolumn{15}{c}{ }    \\
(5) & -1.5 (16.7) & 18.0 & 96.2 & 26.0 & 98.7 & -0.7 (12.5) & 13.2 & 95.1 & 14.0 & 96.3 & -0.7 (11.4) & 11.7 & 95.9 & 11.7 & 95.9 \\
(6) & -0.2 (17.5) & 18.6 & 96.3 & 25.7 & 98.9 & -0.2 (13.3) & 13.7 & 94.9 & 14.4 & 95.6 & -0.3 (11.6) & 11.8 & 94.9 & 11.9 & 94.6  \\
	\hline			
	 &    \multicolumn{15}{c}{ 	\multirow{2}{*}{Doubly Robust Estimator ($\hat \tau_{dr}$)} }    \\
	 &    \multicolumn{15}{c}{ }    \\
(5) &  -2.0 (15.3) & 21.3 & 98.3 & 15.3 & 94.7 & -1.0 (12.6) & 16.3 & 98.3 & 13.2 & 95.7 & -0.9 (11.8) & 13.1 & 97.2 & 12.0 & 95.5 \\
(6) & -0.8 (16.1) & 22.1 & 98.7 & 16.0 & 94.9 & -0.6 (13.4) & 16.8 & 98.2 & 13.5 & 94.2 & -0.4 (11.7) & 13.4 & 97.4 & 12.1 & 94.5   \\
  \hline
 		 &   \multicolumn{15}{c}{ 	\multirow{2}{*}{\cite{athey2016estimating}'s Method} }    \\
	 &    \multicolumn{15}{c}{ }    \\
(5) & -97.8 (127.6) & 128.5 & 88.5 & 124.8 & 85.7 & -94.7 (88.5) & 90.3 & 83.0 & 89.0 & 81.5 & -97.4 (60.8) & 63.5 & 66.9 & 62.9 & 65.8  \\
(6) & -89.8 (128.9) & 131.8 & 89.4 & 127.6 & 88.2 & -93.7 (92.7) & 92.3 & 81.9 & 90.6 & 81.0 & -99.4 (65.9) & 65.1 & 67.9 & 64.3 & 66.7  \\ 
	\hline 	\hline
\end{tabular}}
%\begin{flushleft}
%{\footnotesize Note: All the values in this table have been magnified 100 times. Bias and SD are the Monte Carlo bias and   
%standard deviation over the 1000 simulations of the points estimates. 
% ESE and CP95 are the averages of estimated asymptotic standard error and coverage proportions of the 95\% confidence intervals based on the plug-in method, respectively. ESE.b and CP95.b have the same meaning as ESE and CP95 but are derived from 50 bootstraps.}
%\end{flushleft}
\label{tab-r2}
\end{table}

  % ========================
  
\bigskip \noindent 
{\bf Binary outcome.}  Corresponding to cases (1)-(6), we set 6 simulations (cases (7)-(12)) to 
   evaluate the performance of the proposed methods with binary outcomes.  The data generation mechanisms for cases (7)-(12) are provided in Section 5.1 of Supplementary Material. 
   Tables \ref{tab-r3}-\ref{tab-r4} summarize the numeric results for cases (7)-(12). 
   
 \begin{table}
\centering
\caption{Comparison of various estimators for cases (7)-(10),  binary outcome.  The distributions of $U$ between the RCT and observational data are the same for cases (7)-(8) and different for cases (9)-(10).} 
\resizebox{\linewidth}{!}{\begin{tabular}{c rcccc  rcccc rcccc } 
\hline  	\hline
	 &    \multicolumn{5}{c}{$n_1 =50$}    & \multicolumn{5}{c}{$n_1 =100$} &  \multicolumn{5}{c}{$n_1 =200$}  \\
Case    & Bias (SD) & ESE & CP95 & ESE.b & CP95.b  & Bias (SD) & ESE & CP95 & ESE.b & CP95.b    & Bias (SD) & ESE & CP95 & ESE.b & CP95.b  \\  \hline  
			 &   \multicolumn{15}{c}{ 	\multirow{2}{*}{IPW Estimator ($\hat \tau_{ipw}$), with True Propensity Score} }    \\
	 &    \multicolumn{15}{c}{ }    \\
  (7)  &  0.0 (20.7) & 20.3 & 93.7 & 20.2 & 92.3 & 0.0 (14.7) & 14.7 & 95.0 & 14.7 & 94.0 & -0.4 (10.7) & 10.6 & 94.2 & 10.7 & 93.8   \\
  (8) &   0.1 (25.0) & 25.6 & 95.2 & 25.3 & 94.3 & 0.4 (18.2) & 18.3 & 95.2 & 18.3 & 94.6 & -0.2 (12.9) & 13.3 & 95.5 & 13.3 & 95.0 \\
  (9) &  0.9 (19.6) & 20.5 & 95.0 & 20.3 & 94.4 & 0.4 (15.0) & 14.8 & 93.9 & 14.8 & 93.9 & 0.0 (10.8) & 10.8 & 94.8 & 10.9 & 94.6    \\
  (10)  & -0.2 (24.5) & 27.3 & 95.9 & 25.7 & 94.7 & 1.6 (18.8) & 18.7 & 94.5 & 18.7 & 93.4 & 0.5 (14.3) & 13.8 & 93.3 & 13.8 & 92.9   \\
	\hline  
	 &    \multicolumn{15}{c}{ 	\multirow{2}{*}{IPW Estimator ($\tilde \tau_{ipw}$), with Estimated Propensity Score} }    \\
	 &    \multicolumn{15}{c}{ }    \\
  (7)  & 0.0 (8.0) & 7.9 & 94.3 & 9.7 & 96.9 & 0.0 (6.3) & 6.0 & 94.0 & 6.4 & 94.8 & -0.1 (5.0) & 4.9 & 93.6 & 5.2 & 94.4     \\
  (8) & 0.1 (9.6) & 8.5 & 95.8 & 11.2 & 98.5 & 0.8 (6.1) & 6.1 & 94.5 & 6.7 & 95.9 & 0.7 (5.2) & 5.1 & 94.4 & 5.3 & 94.5   \\
  (9) & 0.0 (8.2) & 8.3 & 95.1 & 10.3 & 97.5 & 0.4 (6.3) & 6.3 & 94.8 & 6.7 & 95.6 & 0.1 (5.4) & 5.3 & 93.7 & 5.5 & 94.7     \\
  (10)  & 0.3 (8.8) & 10.4 & 95.1 & 12.0 & 98.1 & 0.7 (7.4) & 6.9 & 92.4 & 7.6 & 94.5 & 0.5 (6.5) & 6.1 & 92.6 & 6.5 & 93.8   \\ 
	\hline			
	 &    \multicolumn{15}{c}{ 	\multirow{2}{*}{Doubly Robust Estimator ($\hat \tau_{dr}$)} }    \\
	 &    \multicolumn{15}{c}{ }    \\
  (7)  &  -0.3 (7.8) & 7.1 & 93.1 & 8.3 & 95.6 & -0.1 (6.5) & 5.2 & 88.4 & 6.7 & 95.0 & -0.1 (5.3) & 3.9 & 86.4 & 5.5 & 94.8   \\
  (8) & 0.2 (7.5) & 7.6 & 95.0 & 8.3 & 95.9 & 0.4 (6.3) & 6.0 & 94.5 & 6.9 & 95.9 & 0.4 (5.6) & 5.0 & 92.8 & 5.8 & 95.1  \\
  (9) &   0.0 (8.1) & 7.1 & 92.1 & 8.7 & 95.2 & 0.2 (6.5) & 5.3 & 89.1 & 7.2 & 95.4 & 0.0 (5.8) & 4.1 & 82.8 & 6.0 & 94.7   \\
  (10)  & -0.5 (9.3) & 8.6 & 94.5 & 9.9 & 96.4 & -0.3 (8.3) & 7.0 & 90.8 & 8.1 & 94.1 & -0.2 (7.2) & 5.9 & 89.3 & 7.2 & 93.1     \\ 
	\hline
 		 &   \multicolumn{15}{c}{ 	\multirow{2}{*}{\cite{athey2016estimating}'s Method} }    \\
	 &    \multicolumn{15}{c}{ }    \\
	   (7)  &   -4.4 (20.5) & 20.2 & 94.1 & 19.8 & 92.4 & -4.7 (14.2) & 14.2 & 93.5 & 14.2 & 92.7 & -4.9 (10.1) & 10.0 & 91.5 & 10.0 & 91.0  \\
  (8) & -6.0 (24.3) & 25.3 & 95.1 & 24.7 & 93.7 & -5.8 (17.4) & 17.8 & 94.2 & 17.6 & 93.9 & -6.5 (12.4) & 12.6 & 92.2 & 12.5 & 91.8 \\
  (9) & -4.0 (19.4) & 20.3 & 95.4 & 19.9 & 93.9 & -4.5 (14.6) & 14.2 & 94.0 & 14.1 & 92.8 & -4.7 (9.9) & 10.0 & 92.7 & 10.0 & 92.0    \\
  (10)  & -6.1 (23.8) & 25.1 & 95.7 & 24.8 & 94.4 & -4.6 (18.0) & 17.8 & 93.4 & 17.8 & 92.3 & -5.8 (13.2) & 13.0 & 91.5 & 12.7 & 90.5 \\
	\hline	 	\hline	
\end{tabular}}
%\begin{flushleft}
%{\footnotesize Note: All the values in this table have been magnified 100 times. Bias and SD are the Monte Carlo bias and   
%standard deviation over the 1000 simulations of the points estimates. 
% ESE and CP95 are the averages of estimated asymptotic standard error and coverage proportions of the 95\% confidence intervals based on the plug-in method, respectively. ESE.b and CP95.b have the same meaning as ESE and CP95 but are derived from 50 bootstraps.}
%\end{flushleft}
\label{tab-r3}
\end{table}

\begin{table}
\centering
\caption{Comparison of various estimators for cases (11)-(12),  binary outcome, discrete $X$ and $S$.}  
\resizebox{\linewidth}{!}{\begin{tabular}{c rcccc  rcccc rcccc } 
\hline 	\hline
	 &    \multicolumn{5}{c}{$n_1 =50$}    & \multicolumn{5}{c}{$n_1 =100$} &  \multicolumn{5}{c}{$n_1 =200$}  \\
Case    & Bias (SD) & ESE & CP95 & ESE.b & CP95.b  & Bias (SD) & ESE & CP95 & ESE.b & CP95.b    & Bias (SD) & ESE & CP95 & ESE.b & CP95.b  \\  \hline  
			 &   \multicolumn{15}{c}{ 	\multirow{2}{*}{IPW Estimator ($\hat \tau_{ipw}$), with True Propensity Score} }    \\
	 &    \multicolumn{15}{c}{ }    \\
(11) &  0.3 (26.8) & 26.5 & 93.5 & 26.1 & 92.8 & 0.7 (18.7) & 18.7 & 94.9 & 18.5 & 93.8 & 0.3 (13.1) & 13.3 & 94.7 & 13.1 & 94.6  \\
(12) &  -0.3 (25.7) & 26.7 & 95.8 & 26.3 & 95.0 & 0.1 (18.7) & 18.9 & 94.9 & 18.7 & 94.2 & -0.2 (14.2) & 13.4 & 93.2 & 13.3 & 93.2  \\ 
	\hline
	 &    \multicolumn{15}{c}{ 	\multirow{2}{*}{IPW Estimator ($\tilde \tau_{ipw}$), with Estimated Propensity Score} }    \\
	 &    \multicolumn{15}{c}{ }    \\
(11) &   0.0 (3.4) & 3.7 & 97.0 & 5.7 & 99.4 & 0.0 (2.5) & 2.6 & 95.0 & 2.8 & 96.2 & 0.1 (2.1) & 2.1 & 96.0 & 2.1 & 95.8 \\
(12) & 0.1 (3.5) & 3.5 & 97.0 & 5.4 & 99.7 & 0.0 (2.4) & 2.4 & 95.5 & 2.6 & 96.4 & 0.1 (1.9) & 2.0 & 95.4 & 2.0 & 95.4 \\
	\hline			
	 &    \multicolumn{15}{c}{ 	\multirow{2}{*}{Doubly Robust Estimator ($\hat \tau_{dr}$)} }    \\
	 &    \multicolumn{15}{c}{ }    \\
(11) &  -0.1 (3.2) & 3.3 & 94.6 & 3.2 & 93.1 & 0.0 (2.5) & 2.7 & 95.7 & 2.6 & 95.4 & 0.0 (2.2) & 2.3 & 95.9 & 2.2 & 94.5  \\
(12) & 0.1 (2.9) & 3.1 & 95.6 & 2.9 & 93.7 & 0.0 (2.4) & 2.5 & 95.1 & 2.4 & 93.2 & 0.1 (2.0) & 2.2 & 96.8 & 2.1 & 94.8  \\ 
  \hline
 		 &   \multicolumn{15}{c}{ 	\multirow{2}{*}{\cite{athey2016estimating}'s Method} }    \\
	 &    \multicolumn{15}{c}{ }    \\
(11) &  -4.5 (26.6) & 27.3 & 94.8 & 26.0 & 93.0 & -4.2 (18.5) & 19.0 & 94.9 & 18.4 & 93.8 & -4.6 (13.0) & 13.3 & 94.4 & 13.1 & 93.9   \\
(12) & -4.9 (25.8) & 27.4 & 96.0 & 26.3 & 94.9 & -4.5 (18.7) & 19.1 & 94.6 & 18.7 & 92.9 & -4.8 (14.1) & 13.4 & 92.3 & 13.2 & 91.2 \\  
	\hline	\hline
\end{tabular}}
%\begin{flushleft}  
%{\footnotesize Note: All the values in this table have been magnified 100 times. Bias and SD are the Monte Carlo bias and   
%standard deviation over the 1000 simulations of the points estimates. 
% ESE and CP95 are the averages of estimated asymptotic standard error and coverage proportions of the 95\% confidence intervals based on the plug-in method, respectively. ESE.b and CP95.b have the same meaning as ESE and CP95 but are derived from 50 bootstraps.}
%\end{flushleft}
\label{tab-r4}
\end{table}

The results presented in Table \ref{tab-r3} are similar to those in Table \ref{tab-r1}, other than 
 the CP95 of the doubly robust estimator is significantly lower than 0.95 for all cases (7)-(10). In comparison, 
the CP95 of IPW estimators $\hat \tau_{ipw}$ and $\tilde \tau_{ipw}$  still have good performance. This indicates the asymptotic variance estimation of  IPW estimators based on the plug-in method is more robust than that of the doubly robust estimator.  
   A possible reason is that the estimation of the doubly robust estimator relies on many parametric model specifications and the asymptotic variance formula based on the plug-in method is valid only when all the models are correctly specified.   
        Nonetheless, the CP95.b of both the doubly robust estimator and IPW estimators performs well in all cases (1)-(12), which means that bootstrap can produce more robust variance estimates than the plug-in method, at a high computational cost.  
 Table \ref{tab-r4} shows the results for discrete $X$ and $S$, again demonstrating the equivalence of the IPW estimator with estimated propensity score and the doubly robust estimator.  
  
  We further explore the finite sample behaviors of the proposed estimators in the scenarios where the unmeasured confounder $U$ affects both $S$ and $Y$ in observational data. In this case, Assumption 3 may not be satisfied.  The corresponding numeric results are similar to those in Tables \ref{tab-r1} and \ref{tab-r3} and are presented in Tables S1 and S2  of Supplementary Material.

In summary,  the simulation results reveal the following phenomena: (1) the IPW estimators ($\hat \tau_{ipw}$ and $\tilde \tau_{ipw}$) have more stable performance than the other two estimators, a possible reason is that IPW estimators rely on parsimonious model specifications;  (2) using the estimated propensity scores can significantly improve the efficiency of IPW estimator; (3) the doubly robust estimator has similar performance to IPW estimator with estimated propensity score concerning Bias and SD.  (4) the method of \cite{athey2016estimating} is less attractive in terms of both Bias and SD.

% ================================================================

\section{Real data analysis}
\label{application}

Immunoglobulin A  nephropathy (IgAN), also called Berger disease, is the most prevalent chronic and primary glomerular disease worldwide \citep{Haas1997, d1987commonest}. The renin-angiotensin-aldosterone system (RAAS) inhibition is a standard therapy for IgAN disease by slowing proteinuria and lowing blood pressure \citep{Zhang-etal2021}.  
	Despite the usage of RAAS, lgAN patients are still at risk of renal failure \citep{liu2019effects}. Hydroxychloroquine (HCQ), an immunomodulator, is a current therapeutic option for lgAN.  
	Evidence suggests that combination therapy with HCQ and RAAS is effective in reducing proteinuria in patients with IgAN compared to RAAS alone over 6 months \citep{Yang-etal2018}.  However, the long-term effect of HCQ on renal outcomes is less clear \citep{Zhang-etal2021}.  This study aims to  explore the 
	 treatment effect of HCQ on renal failure by combing an RCT dataset and an observational dataset obtained from Peking University First Hospital.

 The RCT data come from a double-blind, randomized, and placebo-controlled trial consisting of 60 observations, of which 30 patients are assigned to the combination therapy with HCQ and RAAS and the rest are assigned to standard RAAS therapy. More details of the RCT data can be found in \citet{liu2019effects}. 
   The observational data contain 547 observations, of which 91 patients accept the combination therapy.  
	 The endpoint (outcome) of interest is a binary variable indicating whether a patient developed renal failure within 3, 4, or 5 years.  In this analysis, we consider two endpoints. \texttt{endpoint 1} is defined as whether glomerular filtration rate (GFR) decreased by 30\%, 40\%, or 50\% from baseline to the end time, \texttt{endpoint 2} is an indicator of whether the GFR is less than 15 ml/min per 1.73 m$^{2}$.   
  Since the randomized controlled trial lasted only six months, no endpoints were observed in RCT data.     
 The surrogate is chosen as the percentage change in proteinuria between baseline and six months. The baseline covariates are the same between the RCT data and observational data, including gender, age, baseline proteinuria, baseline GFR, and some pathologic predictors of renal failure \citep{Shi-etal2011}.  

\begin{table} 
\centering
\caption{Estimated effects of HCQ on renal failure.} 
\resizebox{\linewidth}{!}{\begin{tabular}{cc cc  cc cc } 
\hline
	 &   & \multicolumn{2}{c}{proportion = 0.3}    & \multicolumn{2}{c}{proportion = 0.4} &  \multicolumn{2}{c}{proportion = 0.5}  \\
End time  &   & Endpoint 1 & Endpoint 2 &  Endpoint 1 & Endpoint 2 & Endpoint 1 & Endpoint 2 \\  \hline  
			 &  &   \multicolumn{6}{c}{ 	\multirow{2}{*}{IPW Estimator ($\hat \tau_{ipw}$), with True Propensity Score} }    \\
	 &  &  \multicolumn{6}{c}{ }    \\
 \multirow{2}{*}{3} &  Estimate (ESE.b) & -0.376 (0.11) & -0.202 (0.098) & -0.162 (0.065) & -0.202 (0.103) & -0.123 (0.071) & -0.202 (0.100)    \\
  & $p$-value  &  $<10^{-3}$ & 0.020 & 0.006 & 0.024 & 0.041 & 0.022      \\
 \multirow{2}{*}{4} & Estimate (ESE.b) & -0.418 (0.105) & -0.157 (0.085) & -0.211 (0.073) & -0.157 (0.084) & -0.172 (0.083) & -0.157 (0.077)   \\
  & $p$-value  &  $<10^{-3}$ & 0.033 & 0.002  & 0.032  & 0.019  & 0.021   \\ 
 \multirow{2}{*}{5} &Estimate (ESE.b) & -0.488 (0.109) & -0.192 (0.080) & -0.265 (0.083) & -0.192 (0.080) & -0.211 (0.072) & -0.192 (0.084)  \\
  & $p$-value  &  $<10^{-3}$ & 0.008 & $<10^{-3}$ & 0.008 & 0.002  & 0.011    \\ 
	\hline
	 &  &  \multicolumn{6}{c}{ 	\multirow{2}{*}{IPW Estimator ($\tilde \tau_{ipw}$), with Estimated Propensity Score} }    \\
	 &  &  \multicolumn{6}{c}{ }    \\
 \multirow{2}{*}{3} &  Estimate (ESE.b) &-0.318 (0.082) & -0.178 (0.087) & -0.138 (0.058) & -0.178 (0.089) & -0.102 (0.066) & -0.178 (0.092)   \\
  & $p$-value  &  $<10^{-3}$ & 0.020 & 0.009  & 0.023 & 0.062 & 0.027      \\
 \multirow{2}{*}{4} & Estimate (ESE.b) &  -0.36 (0.079) & -0.145 (0.082) & -0.182 (0.063) & -0.145 (0.089) & -0.155 (0.063) & -0.145 (0.075)   \\
  & $p$-value  &   $<10^{-3}$ & 0.040  & 0.002  & 0.051  & 0.007  & 0.026   \\ 
 \multirow{2}{*}{5} &Estimate (ESE.b) & -0.423 (0.082) & -0.178 (0.082) & -0.232 (0.062) & -0.178 (0.074) & -0.189 (0.065) & -0.178 (0.069) \\
  & $p$-value  &  $<10^{-3}$ & 0.015  & $<10^{-3}$ & 0.008  & 0.002  & 0.005    \\ 
	\hline			
  &  &   \multicolumn{6}{c}{ 	\multirow{2}{*}{Doubly Robust Estimator ($\hat \tau_{dr}$)} }    \\
	 &  &  \multicolumn{6}{c}{ }    \\ 
 \multirow{2}{*}{3} &  Estimate (ESE.b) & -0.256 (0.108) & -0.187 (0.069) & -0.073 (0.102) & -0.187 (0.076) & -0.034 (0.097) & -0.187 (0.073)    \\
  & $p$-value  &  0.009  & 0.004  & 0.237 & 0.007  & 0.362  & 0.005     \\
 \multirow{2}{*}{4} & Estimate (ESE.b) & -0.337 (0.090) & -0.148 (0.065) & -0.125 (0.112) & -0.148 (0.060) & -0.092 (0.099) & -0.148 (0.062)   \\
  & $p$-value  &  $<10^{-3}$ & 0.011 & 0.132 & 0.006 & 0.176  & 0.008    \\ 
 \multirow{2}{*}{5} &Estimate (ESE.b) & -0.396 (0.088) & -0.195 (0.062) & -0.169 (0.116) & -0.195 (0.065) & -0.119 (0.100) & -0.195 (0.064)  \\
  & $p$-value  &  $<10^{-3}$ &  $<10^{-3}$ & 0.074  & 0.001 & 0.118  & 0.001 \\ 
 \hline      
 		 & &   \multicolumn{6}{c}{ 	\multirow{2}{*}{\cite{athey2016estimating}'s Method} }    \\
 	 &  &  \multicolumn{6}{c}{ }    \\
 \multirow{2}{*}{3} &  Estimate (ESE.b) & -0.034 (0.135) & -0.05 (0.084) & -0.045 (0.132) & -0.05 (0.093) & -0.03 (0.082) & -0.05 (0.092)    \\
  & $p$-value  &    0.400 & 0.276 & 0.367 & 0.296  & 0.355 & 0.293    \\
 \multirow{2}{*}{4} & Estimate (ESE.b) &  -0.039 (0.124) & -0.043 (0.071) & -0.071 (0.177) & -0.043 (0.086) & -0.046 (0.100) & -0.043 (0.085)  \\
  & $p$-value  & 0.376 & 0.274  & 0.344 & 0.309  & 0.322 & 0.306    \\ 
 \multirow{2}{*}{5} &Estimate (ESE.b) & -0.052 (0.156) & -0.023 (0.090) & -0.082 (0.15) & -0.023 (0.105) & -0.060 (0.096) & -0.023 (0.085)  \\
  & $p$-value  & 0.368 & 0.397 & 0.292  & 0.412 & 0.266  & 0.392    \\ 
  \hline  
\end{tabular}}
\begin{flushleft}
{\footnotesize Note: ESE.b is estimated asymptotic standard 
error based on 200 bootstraps. The $p$-values are obtained by two-sided test, that is $H_{0}: \tau = 0$ against $H_{1}: \tau \neq 0$}
\end{flushleft}
\label{tab9}
\end{table}

The analysis in \citet{liu2019effects} shows that the new therapy has better efficacy for the surrogate. 
In our analysis, we are interested in estimating the average treatment effect for the long-term outcome in RCT data, thus can determine whether the new therapy has better efficacy than the standard therapy.  
The point estimate and corresponding confidence interval is given in Table~\ref{tab9}. We also test the null hypothesis $H_0: \tau = 0$ versus $ H_1: \tau\neq 0$. As $Y$ is a binary outcome, the regression model $\mu_t(X,S)$ is fitted with logistic regression, and $\mu_t(X)$ is fitted by linear regression of $\mu_t(X,S)$ on $X$. The asymptotic standard errors are obtained based on 200 bootstraps. As shown in Table \ref{tab9},   
IPW method and the doubly robust method have similar point estimates while having different standard errors.  The point estimates of $\tau$ are smaller than zero,  indicating the potential benefit of the new therapy against the standard therapy. The $p$-value for the hypothesis test calculated by IPW and doubly robust methods are smaller than 0.05 in most cases, which means that we can reject the null hypothesis at a significance level of 0.05. Besides, with the end time increasing from 3 years to 5 years,  the absolute value of the point estimate of $\tau$ becomes bigger, which indicates that the efficacy of the new therapy against the standard therapy amplifies over time. 
This result shows a similar pattern that is observed in \citet{liu2019effects}, where the treatment effects on surrogates are analyzed. We also report the results where the asymptotic standard errors are computed with the plug-in method, which are similar to those in Table \ref{tab9} and are presented in Table S3 of Supplementary Material.

% =========================================================

\section{Discussion}
\label{sec7}

This article investigates the average causal effects on the long-term outcome.  Under weaker assumptions than the existing methods,  we derive the semiparametric efficiency bound, propose two new estimators and establish their large sample properties.  Both simulation studies and real data analysis demonstrate the advantages of the proposed method compared with competing ones.  The proposed approach is suitable for various data types of $X$, $S$, and $Y$ and thus has wide application scenarios.

We illustrate the proposed estimators by using generalized linear models to estimate the nuisance parameters. It would be interesting to explore the theoretical properties of the proposed estimators when the nuisance parameters are estimated with machine learning methods \citep{Chernozhukov-etal-2018, Wager-Athey-2018}.   
When $X$ is high-dimensional, one possible extension is to consider how to obtain valid confidence intervals of the proposed estimators when either the propensity score model or the outcome model is correctly specified \citep{vermeulen2015bias, Tan-Annals2020, Sun-Tan2020, Ning-etal2020, Wu-Tan2022}.  
%Another extension is to consider the case where the unmeasured confounders affect the treatment, surrogates, and long-term outcome simultaneously in the observational data. In such a case,  Assumption 3 may not hold.  It is interesting to study how to incorporate sensitivity analysis \citep{Rosenbaum2020} in our proposed method in future research. 
 
%%%%%%%%%%%%%%%%%%%%%%%%%%%%%%%%%%%%%%%%%%%%%%%%%%%%%%%%%%%%%%%%%%%%%%%%%%%%%%%%%%%%%%%%%%%%%%%%%%%%%%%%%%%%%%%%%%%%%%%%%%%%

%%%%%%%%%%%%%%%%%%%%%%%%%%%%%%%%%%%%%%%%%%%%%%%%%%%%%%%%%%%%%%%%%%%%%%%%%%%%%%%%%%%%%%%%%%%%%%%%%%%%%%%%%%%%%%%%%%%%%%%%%%%%
\section*{Supplementary Materials} 
Supplementary Material available online includes  %a heuristic discussion, 
 technical proofs and additional numerical results from the simulation and application.

%%%%%%%%%%%%%%%%%%%%%%%%%%%%%%%%%%%%%%%%%%%%%%%%%%%%%%%%%%%%%%%%%%%%%%%%%%%%%%%%%%%%%%%%%%%%%%%%%%%%%%%%%%%%%%%%%%%%%%%%%%%%
%\section*{Acknowledgments}
%The authors thank the assistant editor and the anonymous reviewers for their helpful comments and valuable suggestions. This research was supported by the State Key Research Program (No. 2021YFF0901400), the National Natural Science Foundation of China (Nos. 11971064, 12071015, and 12171374), and the Major Project of National Statistical Science Foundation of China (No. 2021LD01). 

%%%%%%%%%%%%%%%%%%%%%%%%%%%%%%%%%%%%%%%%%%%%%%%%%%%%%%%%%%%%%%%%%%%%%%%%%%%%%%%%%%%%%%%%%%

%\begin{spacing}{1.25}

\bibhang=1.7pc
\bibsep=2pt
\fontsize{9}{14pt plus.8pt minus .6pt}\selectfont
\renewcommand\bibname{\large \bf References}
%\begin{thebibliography}{11}
\expandafter\ifx\csname
natexlab\endcsname\relax\def\natexlab#1{#1}\fi
\expandafter\ifx\csname url\endcsname\relax
  \def\url#1{\texttt{#1}}\fi
\expandafter\ifx\csname urlprefix\endcsname\relax\def\urlprefix{URL}\fi

%% use bibfile 
  \bibliographystyle{chicago}      % Chicago style, author-year citations
  \bibliography{ref}   % name your BibTeX data base

%\end{spacing}

%%  Another method
\def\thefigure{\arabic{figure}}
\def\thetable{\arabic{table}}

 \renewcommand{\theequation}{S.\arabic{equation}}
 \renewcommand{\thetable}{S\arabic{table}}

\fontsize{12}{14pt plus.8pt minus .6pt}\selectfont \vspace{0.8pc}
\centerline{\large\bf Supplementary Material for ``Identification and estimation}
\vspace{2pt} 
\centerline{\large\bf  of treatment effects on long-term outcomes in clinical trials}
\vspace{2pt} 
\centerline{\large\bf with external observational data"}
\vspace{.4cm} 
\centerline{Wenjie Hu$^{a}$, Xiao-Hua Zhou$^{a, b}$ and Peng Wu$^{c\ast}$\footnote{$\ast$correspond to: pengwu@btbu.edu.cn.}}  
\vspace{.4cm} 
\centerline{$^a$Peking University, $^b$Pazhou Lab, $^c$Beijing Technology and Business University} 
 \vspace{.55cm} \fontsize{9}{11.5pt plus.8pt minus.6pt}\selectfont

\fontsize{12}{14pt plus.8pt minus .6pt}\selectfont

This  Supplementary Material consists of Sections 1--6, where Sections 1-4 give the technical proofs of  Propositions 1--2 and Theorems 1--3, Sections 6 and  7 contain additional numerical results from the simulation study and empirical application. 

% =========================================================

\section{Proof of Propositions 1 and 2}

\emph{Proof of Proposition 1.} 
% \begin{proof} 
Let $e(X) = \pr(T=1\mid X, G=1)$ and \bcol{$h(X, S, T)= E[Y|X, S, T, G=1 ]$}, then we have
% then Proposition 1 follows from the following equations   
\begin{align*}
\tau &= E\left\{\frac{YT}{e(X)} - \frac{Y(1-T)}{1 - e(X)} \mid G=1\right\}  \\
&= E\left\{\frac{h(X,S,T)T}{e(X)} - \frac{h(X,S,T)(1 - T)}{1 - e(X)}\mid G = 1\right\} \\ 
&\bcol{= E\left[E\left\{\frac{h(X,S,T)T}{e(X)} - \frac{h(X,S,T)(1 - T)}{1 - e(X)}\mid X,T,G=1\right\}\mid G = 1\right]}\\
&\bcol{= E \left[ E\left\{h(X,S,T)|X,T=1,G=1  \right\} - E\left\{h(X,S,T)|X,T=0,G=1  \right\} |G=1  \right].}
\end{align*}
\bcol{By Assumption 3, $h(X,S,T)$ can be identified by the $E[Y|X, S, T, G=0]$ in observational data}, so $\tau$ can be identified.

% \end{proof}

\hfill $\Box$

\noindent 
\emph{Proof of Proposition 2.} 
  For $t = 0$ or 1, we have 
\begin{align*}   
& E(Y \mid X, S, T=t, G=1) \\ 
={}& E(Y(t) \mid X, S(t), T=t, G=1)\\
={}& E(Y(t) \mid X, S(t), G = 1)  \qquad \qquad \text{Assumption 1}  \\ 
={}& E(Y(t) \mid X, S(t), G = 0) \qquad \qquad \text{Assumption 4} \\
={}& E(Y(t) \mid X, S(t), T = t, G = 0) \quad ~  \text{Assumption 5} \\  
={}& E(Y|X, S, T=t, G = 0).
\end{align*} 
\hfill $\Box$

% ===================================================================================

% ==========================================================
\section{Proof of Theorem 1}

\emph{Proof of Theorem 1}.  Denote the density of $X$ by $f(x)$, the conditional density $\pr(Y(t) =y|S(t) = s, X=x, G=1)$ by $f_t(y|s,x)$ for $t = 0, 1$, $\pr(S(t) = s|X=x, G=1)$ by $f_t(s|x)$, $\pr(Y=y|S=s, X=x, T=t, G=0)$ by $f(y|s,x,t)$, $\pr(S=s|X = x, T = t, G=0)$ by $f(s|x,t)$, $\pr(T=1|X=x, G=0)$ by $p(x)$ and $\pr(G=1|X=x)$ by $\pi(x)$, then the full observed data distribution is  
\[
\begin{aligned}
f(t,x,s,y,g) &= f(x) \left[ \left\{ f_1(s|x)e(x) \right\}^t \left\{ f_0(s|x)(1 - e(x)) \right\}^{1 - t} \pi(x) \right]^g \\ 
\times& \left[ \left\{ f(y|s,x,1)f(s|x, 1)p(x) \right\}^t \left\{ f(y|s,x,0)f(s|x, 0)(1 - p(x)) \right\}^{1 -t}(1 - \pi(x)) \right]^{1-g}. 
\end{aligned}
\]
% where $f(y|s,x,t) = \pr(Y = y | S = s, X= x, T = t, G = 0) $   and $f(s |x, t) = \pr(S=s | X=x, T=t, G = 0)$ for $t = 0, 1$.

Under Assumptions 1-3 in the manuscript, a parametric submodel indexed by $\theta$ is given as 
\[
\begin{aligned}
&~f(x;\theta) \left[ \left\{ f_1(s|x;\theta)e(x;\theta) \right\}^t \left\{ f_0(s|x;\theta)(1 - e(x;\theta)) \right\}^{1 - t} \pi(x;\theta) \right]^g \\ 
&\times \left[ \left\{ f(y|s,x,1;\theta)f(s|x, 1;\theta)p(x;\theta) \right\}^t \left\{ f(y|s,x,0;\theta)f(s|x, 0;\theta)(1 - p(x;\theta)) \right\}^{1 -t}(1 - \pi(x;\theta)) \right]^{1-g},
\end{aligned}
\]
which equals $f(t,x,s,y,g)$ when $\theta = \theta_0$ and satisfies 
$\int yf_t(y|s,x)dy = \int yf(y|x,s,t)dy$ for $\forall s, x, t$.

The score $S(t,x,s,y,g;\theta)$ is
\[
\begin{aligned}
&S(t,x,s,y,g;\theta)\\ 
=& gtS_1(s|x;\theta)+g(1-t)S_0(s|x;\theta) \\ 
&+ g\frac{t-e(x;\theta)}{e(x;\theta)\{1 - e(x;\theta)\}}\dot{e}(x;\theta) + \frac{g-\pi(x;\theta)}{\pi(x;\theta)\{1 - \pi(x;\theta)\}}\dot{\pi}(x;\theta)\\  
&+ (1-g)t\left\{S(y|s,x,1;\theta) + S(s|x,1;\theta)\right\} + (1-g)(1-t)\left\{S(y|s,x,0;\theta) + S(s|x,0;\theta)\right\} \\ 
&+ (1-g)\frac{t - p(x;\theta)}{p(x;\theta)\{1 - p(x;\theta)\}}\dot{p}(x;\theta) + S_f(x;\theta),
\end{aligned}
\]
where $S_t(s|x;\theta) = \partial \log f_t(s|x;\theta)/\partial \theta$ for $t = 0,1$, $S(y|s,x,t;\theta) = \partial \log f(y|s,x,t;\theta)/\partial \theta$, $S(s|x,t;\theta) = \partial\log f(s|x,t;\theta)/\partial \theta$, and $\dot{e}(x;\theta), \dot{p}(x;\theta), \dot{\pi}(x;\theta)$ are pathwise derivative with respect to $\theta$ for $e(x;\theta), p(x;\theta), \pi(x;\theta)$.

The tangent space $\mathcal{T}$ is 
\[
\begin{aligned}
\mathcal{T} =& \bigg\{ gtS_1(s|x)+ g\{1 - t\}S_0(s|x) + g\{t-e(x)\}a(x) +\{(g-\pi(x)\}b(x)  \\ 
&+  (1-g)t\left\{S(y|s,x,1) + S(s|x,1)\right\} + (1-g)(1-t)\left\{S(y|s,x,0) + S(s|x,0)\right\} \\ 
&+  (1-g)\{t-p(x)\}c(x) + S(x) \bigg\}
\end{aligned}
\]
where $S_t(s|x)$ satisfies $\int S_t(s|x)f_t(s|x)ds = 0$ for $\forall x, t=0,1$, $S(y|s,x,t)$ satisfies  
\newline$\int S(y|s,x,t)f(y|s,x,t)dy = 0$ and $\int yS_t(y|s,x)f_t(y|s,x)dy = \int yS(y|s,x,t)f(y|s,x,t)dy$ for $\forall s, x, t=0,1$, $S(s|x, t)$ satisfies $\int S(s|x,t)f(s|x,t)ds = 0$ for $\forall x, t=0,1$, $S(x)$ satisfies $\int S(x)f(x)dx= 0$, and $a(x), b(x), c(x)$ are arbitrary square-integrable measurable functions.

Under the parametric submodel, the parameter of interest $\tau$ can be represented as 
\[
\begin{aligned}
\tau(\theta) &=  \frac{\int \{\int yf_1(y|s,x;\theta) f_1(s|x;\theta)dyds\} \pi(x;\theta)f(x;\theta)dx }{\int \pi(x;\theta)f(x;\theta)dx} \\ 
& - \frac{\int \{\int yf_0(y|s,x;\theta) f_0(s|x;\theta)dyds\} \pi(x;\theta)f(x;\theta)dx }{\int \pi(x;\theta)f(x;\theta)dx} \\ 
&= \tau_1(\theta) - \tau_0(\theta)
\end{aligned}
\]
where $\tau_1= E(Y(1) |G=1) = \tau_1(\theta_0)$ and $\tau_0=E(Y(0) |G=1) = \tau_0(\theta_0)$ and $\tau = \tau(\theta_0)$. For notation convenience, we denote $\dot{e}(x) = \dot{e}(x;\theta_0), \dot{p}(x) = \dot{p}(x;\theta_0), \dot{\pi}(x) = \dot{\pi}(x;\theta_0)$, $S_f(x) = S_f(x;\theta_0)$, $q = \int \pi(x;\theta)f(x;\theta)dx = \pr(G=1) $ and \newline $\mu_t(x) = E[ Y(t) |X, G= 1] = \int yf_t(y|s,x;\theta) f_t(s|x;\theta)dyds$. Then the pathwise derivative of $\tau_1(\theta)$ is 
\[ 
\begin{aligned}
\left.\frac{\partial\tau_1(\theta)}{\partial\theta}\right|_{\theta = \theta_0}=& \frac{\int \left\{\int y(S_1(y|s,x;\theta_0) + S_1(s|x;\theta_0)) f_1(y|s,x)f_1(s|x)dyds\right\} \pi(x)f(x)dx }{q} \\ 
& + \frac{\int \left\{\mu_1(x) - \tau_1 \right\} \left\{\dot{\pi}(x) + \pi(x)S_f(x)\right\}f(x)dx }{q}\\ 
=& \frac{E\big( \pi(X)E[ E\left\{ Y \cdot S(Y|S,X,1) |S, X, T=1, G=0 \right\} |X, T=1, G=1] \big)}{q} \\ 
&+\frac{ E\left[ \pi(X)E\left\{ \mu_1(S,X) \cdot S_1(S|X)\mid X, T=1, G=1 \right\}  \right] }{q} \\
&+ \frac{E\left[ \{\mu_1(X) - \tau_1\}\{\dot{\pi}(X) + \pi(X)S_f(X)\} \right]}{q}.
\end{aligned}
\]
We let 
\[
\begin{aligned}
\phi_1 =& \frac{G}{q}\left[ \{\mu_1(X) - \tau_1\} + \frac{T\{\mu_1(S,X) - \mu_1(X)\}}{e(X)} \right] \\ 
&+ \frac{1-G}{q}\frac{\pi(X)}{1 - \pi(X)} \frac{T\{Y- \mu_1(S,X)\}}{r(S, X)}\frac{pr(S|X,T=1,G=1)}{pr(S|X,G=0)} \\
=& \frac{G}{q}\left[ \{\mu_1(X) - \tau_1\} + \frac{T\{\mu_1(S,X) - \mu_1(X)\}}{e(X)} \right] \\ 
&+ \frac{1-G}{q}\frac{g_1(S,X) T\{Y - \mu_1(S,X) \}}{e(X)\{1 - g_1(S,X) \}} \\
\end{aligned}
\]
where $\pi(x) =\pr(G=1|X=x)$, $r(s,x) = \pr(T=1|S=s, X=s, G=0)$ and $g_1(s,x) = \pr(G=1|S=s,X=x,T=1)$.

Pathwise differentiability of $\tau_1$ can be verified by 
\begin{equation}\label{pathwise differentiability}
\left.\frac{\partial \tau_1(\theta)}{\partial \theta}\right|_{\theta=\theta_0} = E\left\{ \phi_1 \cdot S(T,X,S,Y,G;\theta_0) \right\}
\end{equation}

Now we give a detailed proof of \eqref{pathwise differentiability}.
\[
\begin{aligned}
&E\left\{\phi_1 \cdot S(T,X,S,Y,G;\theta_0)\right\} \\
=& \frac{1}{q} E \bigg[  G\{\mu_1(X) - \tau_1\}S(T,X,S,Y,G;\theta_0)  \\ 
&+ \frac{GT\{\mu_1(S,X) - \mu_1(X)\}}{e(X)}S(T,X,S,Y,G;\theta_0) \\ 
&+ \frac{(1 - G)\pi(X)}{1 - \pi(X)}\frac{T\{Y - \mu_1(S,X)\}}{r(S,X)}\frac{pr(S|X,T=1,G=1)}{pr(S|X,G=0)}S(T,X,S,Y,G;\theta_0) \bigg{]}, 
\end{aligned}  
\] 
where 
\[
\begin{aligned}
A =& E\left[ G\{\mu_1(X) - \tau_1\}S(T,X,S,Y,G;\theta_0) \right]\\ 
=& E\left\{ G\{\mu_1(X) - \tau_1\} \left[ TS_1(S|X) + \frac{\{T-e(X)\}\dot{e}(X)}{e(X)\{1 - e(X)\}} + \frac{(G - \pi(X))\dot{\pi}(X)}{\pi(X)\{1 - \pi(X)\}} + S_f(X)  \right] \right\}\\ 
=& E\biggl[ \{\mu_1(X) - \tau_1\}\{\dot{\pi}(X) + \pi(X)S_f(X)\} \biggl], 
\end{aligned}
\]

\[
\begin{aligned}
B =& E\left[ \frac{GT(\{\mu_1(S,X) - \mu_1(X)\}}{e(X)}S(T,X,S,Y,G;\theta_0) \right] \\ 
=& E\left[ \frac{GT\{\mu_1(S,X) - \mu_1(X)\}}{e(X)}S_1(S|X) \right] \\ 
=& E\biggl( \pi(X) E\left[ \{\mu_1(S,X) - \mu_1(X)\}S_1(S|X)\mid X, T=1, G=1 \right] \biggl)\\ 
=& E\biggl\{ \pi(X) E\left[ \mu_1(S,X)S_1(S|X)\mid X, T=1, G=1 \right] \biggl\},  \\
C =& E\left[ (1-G)\frac{\pi(X)}{1 - \pi(X)} \frac{T\{Y-\mu_1(S,X)\}}{r(S,X)}\frac{pr(S|X,T=1,G=1)}{pr(S|X,G=0)} S(T,X,S,Y,G;\theta_0) \right] \\ 
=& E\left[ (1 - G)\frac{\pi(X)}{1 - \pi(X)} \frac{T\{Y-\mu_1(S,X)\}}{r(S,X)}\frac{pr(S|X,T=1,G=1)}{pr(S|X,G=0)} S(Y|S,X,1)  \right] \\ 
=& E\biggl( \pi(X) E\biggl\{ E\big[ \{Y - \mu_1(S,X)\}S(Y|S,X,1)\frac{pr(S|X,T=1,G=1)}{pr(S|X,G=0)}\\
&|S, X, T=1, G=0 \big] |X, G=0 \biggl\} \biggl) \\ 
=& E\biggl\{ \pi(X)E( E\left[ Y S(Y|S,X,1)|S, X, T=1, G=0 \right] |X,T=1,G=1) \biggl\}
\end{aligned}
\]

Therefore, we have
\[
\begin{aligned}
& E\left\{ \phi_1 S(T,X,S,Y,G;\theta_0) \right\} \\ 
&= \frac{1}{q}(A+B+C) \\ 
&= \left.\frac{\partial\tau_1(\theta)}{\partial\theta}\right|_{\theta = \theta_0}
\end{aligned}
\]

In addition, $\phi_1$ can be represented as an element in $\mathcal{T}$. Let
\[
\begin{aligned}
S_1(s|x) &= \frac{1}{q} \frac{\mu_1(s,x) - \mu_1(x)}{e(x)}, \\ 
S(y|s,x,1) &= \frac{1}{q}\frac{\pi(x)}{1-\pi(x)}\frac{\{y-\mu_1(s,x)\}}{r(s,x)}\frac{pr(S=s|X=x,T=1,G=1)}{pr(S=s|X=x,G=0)}, \\
a(x) &= c(x)= 0, \\
b(x) &= \frac{1}{q}\{\mu_1(x) - \tau_1\}, \\ 
S(x) &= \frac{1}{q}\pi(x)\{\mu_1(x) - \tau_1\}.
\end{aligned}
\]
Then $\phi_1$ can be decomposed as $\phi_1(y,s,x,t,g) = gtS_1(s|x)+(g-\pi(x))b(x) + (1-g)tS(y|s,x,1) + S(x) $. And the functions satisfy
\[
\begin{aligned}
\int S_1(s|x) f_1(s|x) ds &= 0, \\ 
\int S(y|s,x,1) f(y|s,x,1)dy &= 0, \\ 
\int S(x)f(x)dx &= 0, \\ 
\int yS_1(y|s,x)f_1(y|s,x) dy &= \int yS(y|s,x,1) f(y|s,x,1) dy, \\ 
\end{aligned}
\]
if we choose $S_1(y|s,x) = S(y|s,x,1) f(y|s,x,1)/f_1(y|s,x)$. So $\phi_1$ is in the tangent space. For $\tau_0$, we can get similar procedure and 
\[
\begin{aligned}
\phi_0 =& \frac{G}{q}\left[ \{\mu_0(X) - \tau_0\} + \frac{(1-T)\{\mu_0(S,X) - \mu_0(X)\}}{1 - e(X)} \right] \\
&+ \frac{1-G}{q}\frac{\pi(X)}{1 - \pi(X)} \frac{(1-T)\{Y- \mu_0(S,X)\}}{1-r(S, X)}\frac{pr(S|X,T=0,G=1)}{pr(S|X,G=0)} \\ 
=& \frac{G}{q}\left[ \{\mu_0(X) - \tau_0\} + \frac{(1-T)\{\mu_0(S,X) - \mu_0(X)\}}{1 - e(X)} \right] \\
&+ \frac{1-G}{q}\frac{g_0(S,X)(1-T)\{Y- \mu_0(S,X)\}}{\{1-e(X)\}\{ 1 - g_0(S,X) \}} \\ 
\end{aligned}
\]
where $g_0(s,x) = f(G=1|S=s,X=x,T=0)$.
So we have the efficient influence functions for $\tau$ is 
\[
\begin{aligned}
\phi =& \phi_1 - \phi_0 \\ 
=& \frac{G}{q}\left\{ \frac{T\{\mu_1(S, X) - \mu_1(X)\}}{e(X)} - \frac{(1-T)\{\mu_0(S, X) - \mu_0(X)\}}{1 - e(X)} + \{\mu_1(X) - \mu_0(X) \}- \tau \right\} \\ 
&+ \frac{1-G}{q}\biggl\{\frac{g_1(S,X) T\{Y - \mu_1(S,X) \}}{e(X)\{1 - g_1(S,X) \}} -  \frac{g_0(S,X)(1-T)\{Y- \mu_0(S,X)\}}{\{1-e(X)\}\{ 1 - g_0(S,X) \}} \biggl\}
\end{aligned}
\]

As $\tau$ can be represented as a functional of part of the likelihood, which is irrelevant with  the propensity score models $e(X)$ and $r(S,X)$, so the efficient influence function is the same whether the propensity score models are known or not. 

To show that in the nonparametric model, $\phi$ is the only influence function,  we need only to show that $\mathcal{T}$ contains all the mean zero functions of the observed data. For $h(T,X,S,Y,G)$ that satisfies $E(h(T,X,S,Y,G)) = 0$, we can decompose it as
\[
\begin{aligned}
&h(T,X,S,Y,G) \\
=& GT h_1(X,S) + G(1-T)h_2(X,S) \\ 
&+ (1-G)Th_3(X,S,Y) + (1-G)(1 -T)h_4(X,S,Y) \\ 
=& GT\big[h_1(X,S) - E\{ h_1(X,S)|X \} \big] + G(1 - T) \big[h_2(X,S) - E\{h_2(X,S)|X \} \big] \\
&+G\{T-e(X)\}\big[E\{h_1(X,S)|X \} - E\{h_2(X,S)|X \}\big]   \\ 
&+ (1 - G)T\big[ h_3(X,S,Y) - E\{h_3(X,S,Y)|X,S \} +  E\{h_3(X,S,Y)|X,S \} - E\{h_3(X,S,Y)|X\} \big] \\ 
&+ (1-G)(1-T)\big[h_4(X,S,Y) - E\{h_4(X,S,Y)|X,S \} +  E\{h_4(X,S,Y)|X,S \} - E\{h_4(X,S,Y)|X\} \big] \\ 
&+ (1-G)\{T - p(X) \} \big[E\{h_3(X,S,Y)|X\} - E\{h_4(X,S,Y)|X\} \big] \\ 
&+ \{G - \pi(X) \} \biggl( e(X)\big[E\{h_1(X,S)|X \} - E\{h_2(X,S)|X \}\big] + E\{h_2(X,S)|X \} \\ 
&-  p(X)\big[E\{h_3(X,S,Y)|X\} - E\{h_4(X,S,Y)|X\} \big] - E\{ h_4(X,S,Y)|X \} \biggl) \\
&+\pi(X)\biggl( e(X)\big[E\{h_1(X,S)|X \} - E\{h_2(X,S)|X \}\big] + E\{h_2(X,S)|X \} \\ 
&-  p(X)\big[E\{h_3(X,S,Y)|X\} - E\{h_4(X,S,Y)|X\} \big] - E\{ h_4(X,S,Y)|X \} \biggl)
\end{aligned}
\]
therefore  we let
\[
\begin{aligned}
S_1(S|X) &= h_1(X,S) - E\{ h_1(X,S)|X \} \\ 
S_0(S|X) &= h_2(X,S) - E\{ h_2(X,S)|X \} \\ 
a(X) &= E\{h_1(X,S)|X \} - E\{h_2(X,S)|X \} \\ 
b(X) &= \biggl( e(X)\big[E\{h_1(X,S)|X \} - E\{h_2(X,S)|X \}\big] + E\{h_2(X,S)|X \} \\ 
&-  p(X)\big[E\{h_3(X,S,Y)|X\} - E\{h_4(X,S,Y)|X\} \big] - E\{ h_4(X,S,Y)|X \} \biggl) \\
S(Y|S,X,1) &= h_3(X,S,Y) - E\{h_3(X,S,Y)|X,S \}   \\ 
S(S|X,1)   &= E\{h_3(X,S,Y)|X,S \} - E\{h_3(X,S,Y)|X\} \\ 
S(Y|S,X,0) &= h_4(X,S,Y) - E\{h_4(X,S,Y)|X,S \}   \\ 
S(S|X,0)   &= E\{h_4(X,S,Y)|X,S \} - E\{h_4(X,S,Y)|X\} \\ 
c(X)       &= E\{h_3(X,S,Y)|X\} - E\{h_4(X,S,Y)|X\} \\ 
S(X) &= \pi(X)\biggl( e(X)\big[E\{h_1(X,S)|X \} - E\{h_2(X,S)|X \}\big] + E\{h_2(X,S)|X \} \\ 
&-  p(X)\big[E\{h_3(X,S,Y)|X\} - E\{h_4(X,S,Y)|X\} \big] - E\{ h_4(X,S,Y)|X \} \biggl) 
\end{aligned}
\]
So any function of observed data $h(T,X,S,Y,G)$ that has mean zero can be represented as an element of $\mathcal{T}$, so $\phi$ is the unique influence function for $\tau$ in the nonparametric model.

\hfill $\Box$

% ==============================================
\section{Proof of Theorem 2}

\emph{Proof of Theorem 2}. 
Following the proof of \citet{vermeulen2015bias}, we first give the influence function for $\hat{\tau}_{dr}$, then the double robustness and asymptotically normal follow in a straightforward way.

Denote $Z = (T,X,S,Y,G)$ and
\[
\begin{aligned}
&\psi(Z; p,\alpha, \beta,\gamma, \eta) \\
=&  \frac{G}{q}\bigg[ \frac{T\{\mu_1(S, X;\alpha_1) - \mu_1(X;\beta_1)\}}{e(X;\gamma)} - \frac{(1-T)\{\mu_0(S, X;\alpha_0) - \mu_0(X;\beta_0)\}}{1 - e(X;\gamma)} + \mu_1(X;\beta_1) - \mu_0(X;\beta_0) - \tau \bigg]  \\
&+ \frac{1-G}{q}\biggl[\frac{g_1(S,X;\eta_1) T\{Y - \mu_1(S,X;\alpha_1) \}}{e(X;\gamma)\{1 - g_1(S,X;\eta_1) \}} -  \frac{g_0(S,X;\eta_0)(1-T)\{Y- \mu_0(S,X;\alpha_0)\}}{\{1-e(X;\gamma)\}\{ 1 - g_0(S,X;\eta_0) \}} \biggl],
\end{aligned}
\]
where $\alpha = (\alpha_1^\T, \alpha_0^\T)^\T, \beta = (\beta_1^\T, \beta_0^\T)^\T $ and $\eta = (\eta_1^\T, \eta_0^\T)^\T$. We denote $\bar{\alpha}, \bar{\beta}, \bar{\gamma}, \bar{\eta}$ as the probability limits of estimators $\hat{\alpha}, \hat{\beta}, \hat{\gamma}, \hat{\eta}$. Besides, we denote $\alpha^\ast, \beta^\ast, \gamma^\ast, \eta^\ast$ as the true parameter when the working models are correctly specified, respectively. Under suitable regularity conditions \citep{vermeulen2015bias}, 
we have that 
\[
\begin{aligned}
n^{1/2}(\hat{\tau}_{dr} - \tau) =&  n^{1/2}\hat{E}\{\psi(Z;p,\bar{\alpha}, \bar{\beta}, \bar{\gamma}, \bar{\eta}\}  + E\{\frac{\partial \psi}{\partial p}\}n^{1/2}(\hat{p} - p) \\ 
&+ E\{\frac{\partial \psi}{\partial \alpha}\}n^{1/2}(\hat{\alpha} - \bar{\alpha}) + E\{\frac{\partial \psi}{\partial \beta}\}n^{1/2}(\hat{\beta} - \bar{\beta}) \\
&+ E\{\frac{\partial \psi}{\partial \gamma}\}n^{1/2}(\hat{\gamma} - \bar{\gamma}) + E\{\frac{\partial \psi}{\partial \eta}\}n^{1/2}(\hat{\eta} - \bar{\eta})
+ o_{p}(1),
\end{aligned}
\]

So the consistency of $\hat{\tau}_{dr}$ relies on the mean zero property of $\psi(Z;p,\bar{\alpha}, \bar{\beta}, \bar{\gamma}, \bar{\eta})$.
Notice that 
\[
\begin{aligned}
&E\biggl[\frac{T\{\mu_1(S, X; \alpha^\ast_1) - \mu_1(X; \beta^\ast_1) \}}{e(X; \bar{\gamma})} |X, G = 1 \biggl] \\
=& E\biggl[\frac{T\{\mu_1(S(1), X; \alpha^\ast_1) - \mu_1(X;\beta^\ast_1) \}}{e(X; \bar{\gamma})} |X, G = 1 \biggl] \\ 
=& E[ \mu_1(S(1), X;\alpha^\ast_1) - \mu_1(X; \beta^\ast_1) |X, G = 1 ] = 0 \\
&E\biggl[\frac{g_1(S,X;\bar{\eta}_1)T\{ Y -  \mu_1(S, X;\alpha^\ast_1) \}}{e(X;\bar{\gamma})\{1 - g_1(S,X;\bar{\eta}_1)\}} |S, X, T, G = 0 \biggl]  \\ 
=& \frac{g_1(S,X;\bar{\eta}_1)T E \{Y |S, X, T, G = 0 \}}{e(X;\bar{\gamma})\{1 - g_1(S,X;\bar{\eta}_1)  \}} - \frac{g_1(S,X;\bar{\eta}_1)T\mu_1(S,X; \alpha^\ast_1)}{e(X;\bar{\gamma})\{1 - g_1(S,X;\bar{\eta}_1)  \}} \\ 
=& \frac{g_1(S,X;\bar{\eta}_1)T\mu_1(S,X; \alpha^\ast_1)}{e(X;\bar{\gamma})\{1 - g_1(S,X;\bar{\eta}_1)  \}} - \frac{g_1(S,X;\bar{\eta}_1)T\mu_1(S,X; \alpha^\ast_1)}{e(X;\bar{\gamma})\{1 - g_1(S,X;\bar{\eta}_1)  \}} = 0
\end{aligned}
\]
When the condition $(i)$ holds, that is the working regression models $\mu_i(S,X; \alpha_i)$ and $\mu_i(X;\beta_i)$ are correctly specified, thus $\bar{\alpha}, \bar{\beta}$ is the true parameter $\alpha^\ast, \beta^\ast$. So we have
\[
\begin{aligned}
&E\{\psi(Z;p,\alpha^\ast, \beta^\ast, \bar{\gamma}, \bar{\eta})\}  \\ 
=& E\biggl[ \frac{G}{q}\bigg\{ \frac{T E\{\mu_1(S, X;\alpha^\ast_1) - \mu_1(X;\beta^\ast_1)|T=1,X,G=1 \}}{e(X; \bar{\gamma})} \\
&- \frac{(1-T)E\{\mu_0(S, X;\alpha^\ast_0) - \mu_0(X;\beta^\ast_0)|T=0,X,G=1\}}{1 - e(X; \bar{\gamma})} \\ 
&+ \mu_1(X;\beta^\ast_1) - \mu_0(X;\beta^\ast_0) - \tau\biggl\} \\ 
&+ \frac{1-G}{q}\biggl\{\frac{g_1(S,X;\bar{\eta}_1) T\{Y - \mu_1(S,X;\alpha^*_1) \}}{e(X;\bar{\gamma})\{1 - g_1(S,X;\bar{\eta}_1) \}} \\ 
&-  \frac{g_0(S,X;\bar{\eta}_0)(1-T)\{Y- \mu_0(S,X;\alpha^*_0)\}}{\{1-e(X;\bar{\gamma})\}\{ 1 - g_0(S,X;\bar{\eta}_0) \}} \biggl\} \biggl] \\ 
=& E\{\mu_1(X;\beta^\ast_1) - \mu_0(X;\beta^\ast_0) - \tau \}=0
\end{aligned}
\]

When condition $(ii)$ holds, that is the working regression model $\mu_t(s,x;\alpha_t)$ and propensity score model $e(x;\gamma)$ are correctly specified, thus $\bar{\alpha}, \bar{\gamma}$ are equal to $\alpha^\ast, \gamma^\ast$ respectively.

Then we have
\[
\begin{aligned}
&E\{\psi(Z;p,\alpha^\ast, \bar{\beta}, \gamma^\ast, \bar{\eta})\}  \\ 
=& E\biggl[ \frac{G}{q}\bigg\{ \frac{ E[T\{\mu_1(S(1), X;\alpha^\ast_1) - \mu_1(X;\bar{\beta}_1)\}|X,G=1 ]}{e(X; \gamma^\ast)} \\
&- \frac{E[(1-T)\{\mu_0(S(0), X;\alpha^\ast_0) - \mu_0(X;\bar{\beta}_0)\}|X,G=1]}{1 - e(X; \gamma^\ast)} \\ 
&+ \mu_1(X;\bar{\beta}_1) - \mu_0(X;\bar{\beta}_0) - \tau\biggl\} \\ 
&+ \frac{1-G}{q}\biggl\{\frac{g_1(S,X;\bar{\eta}_1) T\{Y - \mu_1(S,X;\alpha^*_1) \}}{e(X;\gamma^*)\{1 - g_1(S,X;\bar{\eta}_1) \}}  \\ 
&-  \frac{g_0(S,X;\bar{\eta}_0)(1-T)\{Y- \mu_0(S,X;\alpha^*_0)\}}{\{1-e(X;\gamma^*)\}\{ 1 - g_0(S,X;\bar{\eta}_0) \}} \biggl\} \biggl] \\ 
=&E \biggl[ E\{\mu_1(S(1), X;\alpha^\ast_1) - \mu_1(X;\bar{\beta}_1)|X,G=1 \}  \\
&-  E\{\mu_0(S(0), X;\alpha^\ast_0) - \mu_0(X;\bar{\beta}_0)|X,G=1\} \\ 
&+ E\{\mu_1(X;\bar{\beta}_1) - \mu_0(X;\bar{\beta}_0)|X,G=1\} -  \tau\biggl] \\ 
=&E\biggl[ E\{\mu_1(S(1), X;\alpha^\ast_1)|X,G=1 \} -  E\{\mu_0(S(0), X;\alpha^\ast_0)|X,G=1\} - \tau\biggl]= 0
\end{aligned}
\]

Therefore we have proved that $E\{\psi(Z;p,\bar{\alpha}, \bar{\beta}, \bar{\gamma}, \bar{\eta})\} = 0$ under condition $(i)$ or $(ii)$, thus $\hat{\tau}_{dr}$ has the doubly robust property. In particular, when all the working models are correctly specified, we have
\[
E\{\frac{\partial \psi}{\partial p}\} = 
E\{\frac{\partial \psi}{\partial \alpha}\}=
E\{\frac{\partial \psi}{\partial \beta}\} =
E\{\frac{\partial \psi}{\partial \gamma}\} =
E\{\frac{\partial \psi}{\partial \eta}\} =  0
\]
thus the influence function of $\hat{\tau}_{dr}$ is exactly the efficient influence function $\phi$.

\hfill $\Box$

% ==============================================
\section{Proof of Theorem 3} 
\emph{Proof of Theorem 3(i)}.  $\sqrt{n_1} ( \hat \tau_{ipw} - \tau )$ can be decomposed as follows.    
\begin{align*} 
\sqrt{n_1} ( \hat \tau_{ipw} - \tau ) 
={}& \sqrt{n_1} \frac{1}{n_1 }	\sum_{i=1}^{n_1} \biggl \{   \frac{ (T_{i}-e_{i})  \cdot \hat h_i   }{  e_{i} (1 - e_{i}) }  - E[  \frac{ (T_{i}-e_{i})  \cdot  h_i   }{ e_{i} (1 - e_{i}) } | G_{i} = 1 ]    \biggr \} \\
:={}&   U_{1n} + U_{2n}, \\
\end{align*} 
where 
\[
\begin{aligned}
U_{1n} ={}&  \sqrt{\frac{ n_{1} }{ n_{0} }}  \cdot \frac{1}{ n_1 }	\sum_{i=1}^{n_1} \biggl \{   \frac{ (T_{i}-e_{i})   }{  e_{i} (1 - e_{i})  } \cdot \sqrt{n_0} (\hat h_i - h_{i})     \biggr \}        \\
U_{2n} ={}& \sqrt{n_1} \cdot \frac{1}{n_1 }	\sum_{i=1}^{n_1} \biggl \{   \frac{ (T_{i}-e_{i})  \cdot  h_i   }{  e_{i} (1 - e_{i})  } -  E[  \frac{ (T_{i}-e_{i})  \cdot  h_i   }{  e_{i} (1 - e_{i})  } | G_{i} = 1 ]    \biggr \}.  \\	
\end{aligned}
\]

We focus on analyzing $U_{1n}$. By a Taylor expansion, we obtain that for $i = 1, ..., n_1$, 
\begin{equation}  \label{eq-s1}
\sqrt{n_0}  (\hat h_i - h_{i} ) = [h_{i}'(\kappa^*)]^{T} \cdot  \sqrt{n_0} (\hat \kappa -  \kappa^*)  + o_{p}(1).        
\end{equation} 
By the property of maximum likelihood estimation for generalized linear models,    
\begin{equation}   \label{eq-s2}
\sqrt{n_0} (\hat \kappa - \kappa^*)  
 %={}&  \sqrt{n_0}   I^{-1}(\alpha_{0}) ( - \frac{1}{n_0} l'(\alpha_{0}) )  + o_{p}(1) \\
=   I^{-1}(\kappa^*) \frac{1}{ \sqrt{n_0}   } \bcol{\sum_{i=n_1+1}^{n_1+n_0}} (Y_{i} - h_{i}) \tilde X_{i} / \phi + o_{p}(1),  
\end{equation}  
where $I(\kappa^*)$ is the Fisher information matrix of $\kappa$ at $\kappa^*$ in the observational data, $\phi$ is the scale parameter in the corresponding generalized linear model.  Then applying equations (\ref{eq-s1}) and (\ref{eq-s2}) yield that  
\begin{align*}
U_{1n} ={}&  \sqrt{\frac{ n_{1} }{ n_{0} }}  \cdot   \frac{1}{n_1 }	\sum_{i=1}^{n_1} \biggl \{   \frac{ (T_{i}-e_{i})   }{  e_{i} (1 - e_{i}) } \cdot [h_{i}'(\kappa^*)]^{T}  \cdot  \sqrt{n_0} (\hat \kappa - \kappa^*)   \biggr \} + o_{p}(1)            \\
={}&  \sqrt{\frac{ n_{1} }{ n_{0} }}  \cdot   \biggl \{  \frac{1}{n_1 }	\sum_{i=1}^{n_1}   \frac{ (T_{i}-e_{i})   }{  e_{i} (1 - e_{i}) } \cdot h_{i}'(\kappa^*) \biggr \}^{T}  \cdot   I^{-1}(\kappa^*) \frac{1}{ \sqrt{n_0}   } \bcol{\sum_{i=n_1+1}^{n_1+n_0}} (Y_{i} - h_{i}) \tilde X_{i} / \phi  
+ o_{p}(1) \\
={}&  \sqrt{\rho}  \cdot   B_{1}^{T} \cdot   I^{-1}(\kappa^*) \frac{1}{ \sqrt{n_0}   } \bcol{\sum_{i=n_1+1}^{n_1+n_0}} (Y_{i} - h_{i}) \tilde X_{i}  / \phi 
+ o_{p}(1), 
\end{align*}
where 
		\[ B_{1}  =  E \biggl [  \frac{ (T_{i}-e_{i})   }{  e_{i} (1 - e_{i}) } \cdot h_{i}'(\kappa^*) \mid G_{i} = 1 \biggr ].   \]
		Thus, 
\begin{equation}  \label{eq-s3}
\begin{split}  
\sqrt{n_1} ( &   \hat \tau_{ipw} - \tau )  
={}   \sqrt{\rho}  \cdot   B_{1}^{T} \cdot   I^{-1}(\kappa^*) \frac{1}{ \sqrt{n_0}   } \bcol{\sum_{i=n_1+1}^{n_1+n_0}} (Y_{i} - h_{i}) \tilde X_{i}/ \phi  +  \\
 {}& \frac{1}{ \sqrt{n_1} }	\sum_{i=1}^{n_1} \biggl [   \frac{ (T_{i}-e_{i})  \cdot  h_i   }{  e_{i}  (1 - e_{i} ) } -  E \Big \{  \frac{ (T_{i}-e_{i})  \cdot  h_i   }{  e_{i}  (1 - e_{i} ) } | G_{i} = 1  \Big\}    \biggr ] + o_{p}(1).  
 \end{split}
\end{equation}
%U_{1n} + U_{2n} =	
	Since the first two terms on the right-hand side of the above equation are independent, then Theorem 3(i) follows immediately from equation (\ref{eq-s3}) the true that 
 \[
\begin{aligned}     
& var\{  \sqrt{\rho} B_{1}^{T}  \cdot   I^{-1}(\kappa^*) \frac{1}{ \sqrt{n_0}  }\bcol{\sum_{i=n_1+1}^{n_1+n_0}} (Y_{i} - h_{i}) \tilde X_{i} / \phi \}  \\
={}& \rho B_{1}^{T}  \cdot   I^{-1}(\kappa^*)  \cdot var\{(Y_{i} - h_{i}) \tilde X_{i} / \phi\} \}   I^{-1}(\kappa^*)  B_{1}   =  \rho B_{1}^{T} I^{-1}(\kappa^*) B_{1}. 
\end{aligned}
\]
\hfill $\Box$

% 
% By a similar decomposition to $\hat \tau_{ipw}$, 
\bigskip 
\noindent 
\emph{Proof of Theorem 3(ii)}.   We decompose $\sqrt{n_1} ( \tilde \tau_{ipw} - \tau )$ as follows.  
\[
\begin{aligned} 
\sqrt{n_1} ( \tilde \tau_{ipw} - \tau ) 
={}& \sqrt{n_1} \frac{1}{n_1 } \sum_{i=1}^{n_1} \biggl \{   \frac{ (T_{i}- \hat e_{i})  \cdot \hat h_i   }{  \hat e_i (1 - \hat e_i) }  - \tau   \biggr \} \\
={}& \sqrt{n_1} \frac{1}{n_1 } \sum_{i=1}^{n_1} \biggl \{  \frac{ (T_{i}- \hat e_{i})  \cdot \hat h_i   }{  \hat e_i (1 - \hat e_i) } - \frac{ (T_{i}-  e_{i})  \cdot \hat  h_i   }{  e_i (1 - e_i) } + \frac{ (T_{i}-  e_{i})  \cdot \hat  h_i   }{  e_i (1 - e_i) }  - \tau   \biggr \} \\   
:={}&   U_{3n} - U_{4n} + 
\sqrt{n_1} ( \hat \tau_{ipw} - \tau ), \\
\end{aligned} 
\]
where 
    \begin{align*}
        U_{3n} ={}& \sqrt{n_1} \frac{1}{n_1 } \sum_{i=1}^{n_1} \Big \{ \frac{T_i \hat h_i}{\hat e_i} - \frac{T_i \hat h_i}{ e_i}   \Big \}, \\
        U_{4n} ={}& \sqrt{n_1} \frac{1}{n_1 } \sum_{i=1}^{n_1} \Big \{ \frac{ (1 - T_i) \hat h_i}{ 1- \hat e_i} - \frac{(1-T_i) \hat h_i}{ 1 - e_i}   \Big \}, 
    \end{align*}
    
Firstly, we discuss $U_{3n}$.  According to the properties of logistic regression, 
    \begin{equation}   \label{eq-s4}  
    \begin{split}
         \sqrt{n_1}(\hat e_i - e_i) 
         ={}& e_i (1 - e_i) X_i^T \cdot \sqrt{n_1} (\hat \gamma - \gamma^*) + o_p(1),    \\
         ={}& e_i (1 - e_i) X_i^T \cdot I^{-1}(\gamma^*) \cdot \frac{1}{\sqrt{n_1}} \sum_{j=1}^{n_1} X_j (T_j - e_j) + o_p(1),    
         \end{split}
    \end{equation} 
where   $I(\gamma^*)$ is the Fisher information matrix of $\gamma$ at $\gamma^*$.  By equation (\ref{eq-s4}),  
    \begin{align}
        U_{3n} ={}& -  \sqrt{n_1} \frac{1}{n_1 } \sum_{i=1}^{n_1} \frac{ T_i \hat h_i (\hat e_i - e_i) }{ e_i^2 } + o_p(1) \notag \\
        ={}& - \sqrt{n_1} \frac{1}{n_1 } \sum_{i=1}^{n_1} \frac{ T_i  h_i (\hat e_i - e_i) }{ e_i^2 } + o_p(1)  \notag \\
        ={}& -  \Big \{ \frac{1}{n_1} \sum_{i=1}^{n_1} \frac{ T_i  h_i }{ e_i^2 } e_i (1 - e_i) X_i \Big \}^T \cdot I^{-1}(\gamma^*) \cdot \frac{1}{\sqrt{n_1}} \sum_{j=1}^{n_1} X_j (T_j - e_j) + o_p(1).  \label{eq-s5}
     %   ={}& -   A_2^T I^{-1}(\beta_0) \cdot n_1^{-1/2} \sum_{j=1}^{n_1} X_j (T_j - e_j) + o_p(1), 
    \end{align}

Secondly, we consider $U_{4n}$. Through a similar argument to $U_{3n}$,  
    \begin{align} 
        U_{4n} ={}&  \sqrt{n_1} \frac{1}{n_1 } \sum_{i=1}^{n_1} \frac{ (1-T_i)  h_i (\hat e_i - e_i) }{ (1-e_i)^2 } + o_p(1) \notag \\
        ={}&    \Big \{ \frac{1}{n_1} \sum_{i=1}^{n_1} \frac{ (1-T_i)  h_i }{ (1-e_i)^2 } e_i (1 - e_i) X_i \Big \}^T \cdot I^{-1}(\gamma^*) \cdot \frac{1}{\sqrt{n_1}} \sum_{j=1}^{n_1} X_j (T_j - e_j) + o_p(1).  \label{eq-s6}
    \end{align}

Finally, let 
		\[    B_{2} =   E[ \frac{ T_i h_i (1-e_i)X_i }{e_i} | G_i = 1 ] +     E[ \frac{ (1-T_i) h_i e_i X_i }{ 1 - e_i } | G_i = 1 ].    \]
Combing (\ref{eq-s3}), (\ref{eq-s5}) and (\ref{eq-s6}) yield that  
    \begin{align*}
        \sqrt{n_1} ( \tilde \tau_{ipw} - \tau ) 
={}& - B_{2}^T \cdot I^{-1}(\gamma^*) \cdot \frac{1}{\sqrt{n_1}}  \sum_{i=1}^{n_1} X_i (T_i - e_i) + \sqrt{n_1} ( \hat \tau_{ipw} - \tau ) + o_p(1)  \\ 
={}& -  B_{2}^T \cdot I^{-1}(\gamma^*) \cdot \frac{1}{\sqrt{n_1}}  \sum_{i=1}^{n_1} X_i (T_i - e_i) +  \\ 
{}&   \sqrt{\rho}  \cdot   B_{1}^{T} \cdot   I^{-1}(\kappa^*) \frac{1}{ \sqrt{n_0}   } \bcol{\sum_{i=n_1+1}^{n_1+n_0}} (Y_{i} - h_{i}) \tilde X_{i}/ \phi +  \\
 {}& \frac{1}{ \sqrt{n_1} }	\sum_{i=1}^{n_1} \biggl [   \frac{ (T_{i}-e_{i})  \cdot  h_i   }{  e_{i}  (1 - e_{i} ) } -  E \Big \{  \frac{ (T_{i}-e_{i})  \cdot  h_i   }{  e_{i}  (1 - e_{i} ) } | G_{i} = 1  \Big\}    \biggr ]  + o_{p}(1). 
    \end{align*} 
Note that 
	\[  var\left\{  -  B_{2}^T \cdot I^{-1}(\gamma^*) \cdot \frac{1}{\sqrt{n_1}}  \sum_{i=1}^{n_1} X_i (T_i - e_i)   \right\}  = B_{2}^{T} \cdot   I^{-1}(\gamma^*)  \cdot B_{2},  \]
 and 	
    \begin{align*}
        cov \left \{ X_i(T_i - e_i),   \frac{ (T_{i}-e_{i})  \cdot  h_i   }{  e_{i}  (1 - e_{i} ) }  \right  \}
        ={}& cov \left \{ X_i(T_i - e_i),  \frac{T_ih_i}{e_i} -\frac{(1-T_i)h_i}{1 - e_i}    \right  \}
         \\
        ={}& 
        E[ X_i(T_i - e_i) \frac{T_ih_i}{e_i} ] -  E[ X_i(T_i - e_i) \frac{(1-T_i)h_i}{1 - e_i} ]     \\
        ={}& B_{2},
    \end{align*}   
 then the variance of $\sqrt{n_1} ( \tilde \tau_{ipw} - \tau )$ is 
     	\begin{align*}
   var\{  \sqrt{n_1} ( \tilde \tau_{ipw} - \tau ) \}	
   ={}&   B_{2}^{T} \cdot   I^{-1}(\gamma^*)  \cdot B_{2} +  var\{  \sqrt{n_1} ( \hat \tau_{ipw} - \tau )  \}  \\
     {}&  -  2   B_{2}^T \cdot I^{-1}(\gamma^*) \cdot       cov \left \{ X_i(T_i - e_i),   \frac{ (T_{i}-e_{i})  \cdot  h_i   }{  e_{i}  (1 - e_{i} ) }  \right  \}  + o_{p}(1) \\
     ={}&   B_{2}^{T} \cdot   I^{-1}(\gamma^*)  \cdot B_{2} +  var\{  \sqrt{n_1} ( \hat \tau_{ipw} - \tau )  \}   -  2   B_{2}^T \cdot I^{-1}(\gamma^*) \cdot   B_{2} +  o_{p}(1) \\
     ={}&   var\{  \sqrt{n_1} ( \hat \tau_{ipw} - \tau )  \}   -  B_{2}^T \cdot I^{-1}(\gamma^*) \cdot   B_{2} +  o_{p}(1). 
	\end{align*}
	This completes the proof of Theorem 3(ii). 
	
\hfill $\Box$

% ==============================================
\section{Additional Results for Simulation Study}

\subsection{Data generation mechanisms for binary outcome}
  
    In contrast with cases (1)-(6), the data generating process of cases (7)-(12) are as follows: 
     
     {\bf Case (7)}.  For RCT data ,  $S = U + 2(X_{1}+X_{2}) + T + \varepsilon_S$,  $\pr(Y=1 \mid T, X, S, U)  = \text{expit}\{ T + 3(X_{1} + X_{2}) + S \}$, $X = (X_{1}, X_{2})^{T} \sim N(0,I_2)$.   
       For observational data,  $X \sim N(1, 4 I_2), U \sim N(0,1)$,  $\pr(T=1| X, U) = \text{expit}\{ U + X_{1} + X_{2} \}$, $S$ and $Y$ are generated the same as in RCT data . 
     
    {\bf Case (8)}. For RCT data , $S = U^2 + 2(X_{1}^2 + X_{2}^{2})  + T + \varepsilon_S,$  $\pr(Y=1 \mid T, X, S, U)  = \text{expit}\{ T + 3(X_{1} + X_{2}) + S \}$, $X = (X_{1}, X_{2})^{T} \sim N(0, I_2)$.  
      For observational data,  $X\sim N(0, 4 I_2), U \sim N(0,1)$,   $\pr(T=1| X, U) = \text{expit}\{ U + X_{1} + X_{2} \}$, $S$ and $Y$ are generated the same as in RCT data . 
     
{\bf Case (9)}.  The data generation mechanism is the same as in case (7), except for setting $U \sim N(1, 4)$ in observational data. 
     
    {\bf Case (10)}.  The data generation mechanism is the same as in case (8), except for setting  $U \sim N(1, 4)$ in observational data. 
    
      {\bf Cases (11) and (12)}.  The data generating process of cases (11)-(12) are the same as cases (7)-(8), respectively, except that setting $X$ and $S$ as discrete variables.   
     For cases (11)-(12), the covariates $X$ consisting of two binary variables $(X_{1}, X_{2})$, $X_{1}$ and $X_{2}$ are independent and identically distributed from a Binomial distribution $B(1, 0.5)$ for both the RCT and observational data. The surrogate $S$ is generated through a logistic regression with $\pr(S = 1 \mid  X, S, U, T) = \text{expit}\{U + 2(X_{1}+X_{2}) + T\}$ for case (11) and $\pr(S = 1 \mid  X, S, U, T) = \text{expit}\{U^2 + 2(X_{1}^2 + X_{2}^{2})  + T\}$ for case (12).

\subsection{Additional simulation: Assumption Violation}  

We further consider scenarios where the unmeasured confounder $U$ affects both $S$ and $Y$ in observational data. In this case, Assumption 3 may not be satisfied.  
Four extra cases are set as follows: 

      	{\bf Case (13)}.  For RCT data ,  $S = U + 2(X_{1}+X_{2}) + T + \varepsilon_S$,  $ Y  = U +  T + 3(X_{1} + X_{2}) + S + \varepsilon_Y$, $X = (X_{1}, X_{2})^{T} \sim N(0,I_2)$.   
       For observational data,  $X \sim N(1, 4 I_2), U \sim N(0,1)$,  $\pr(T=1| X, U) = \text{expit}\{ X_{1} + X_{2} \}$, $S$ and $Y$ are generated the same as in RCT data . 
     
    {\bf Case (14)}. For RCT data , $S = U^2 + 2(X_{1}^2 + X_{2}^{2})  + T + \varepsilon_S,$  $Y  =  T + 3(X_{1} + X_{2}) + S + \varepsilon_Y$, $X = (X_{1}, X_{2})^{T} \sim N(0, I_2)$.  
      For observational data,  $X\sim N(1, 4 I_2), U \sim N(0,1)$,   $\pr(T=1| X, U) = \text{expit}\{ X_{1} + X_{2}\}$, $S$ and $Y$ are generated the same as in RCT data. 

 	{\bf Case (15)}.  For RCT data ,  $S = U + 2(X_{1}+X_{2}) + T + \varepsilon_S$,  $\pr(Y=1\mid T, X, S,U)  = \text{expit}\{ U +  T + 3(X_{1} + X_{2}) + S  \}$, $X = (X_{1}, X_{2})^{T} \sim N(0,I_2)$.   
       For observational data,  $X \sim N(1, 4 I_2), U \sim N(0,1)$,  $\pr(T=1| X, U) = \text{expit}\{ X_{1} + X_{2} \}$, $S$ and $Y$ are generated the same as in RCT data . 
     
    {\bf Case (16)}. For RCT data , $S = U^2 + 2(X_{1}^2 + X_{2}^{2})  + T + \varepsilon_S,$  $\pr(Y=1\mid T, X, S,U)  = \text{expit}\{ U +  T + 3(X_{1} + X_{2}) + S \}$, $X = (X_{1}, X_{2})^{T} \sim N(0, I_2)$.  
      For observational data,  $X\sim N(1, 4 I_2), U \sim N(0,1)$,   $\pr(T=1| X, U) = \text{expit}\{ X_{1} + X_{2}\}$, $S$ and $Y$ are generated the same as in RCT data. 

The cases (13)-(14) for continuous outcomes, cases (15)-(16) for binary outcomes. Tables \ref{tab-s1} and \ref{tab-s2} summarize the numeric results for cases (13)-(14). These results are similar to those in  Tables 1 and 3 of the manuscript and indicate that the proposed methods also perform well under the scenario where $U$ affects both $S$ and $Y$.

\begin{table}
\centering
\caption{Comparison of various estimators for cases (13)-(14),  continuous outcome. }  
\resizebox{\linewidth}{!}{\begin{tabular}{c rcccc  rcccc rcccc } 
\hline 
	 &    \multicolumn{5}{c}{$n_1 =50$}    & \multicolumn{5}{c}{$n_1 =100$} &  \multicolumn{5}{c}{$n_1 =200$}  \\
Case    & Bias (SD) & ESE & CP95 & ESE.b & CP95.b  & Bias (SD) & ESE & CP95 & ESE.b & CP95.b    & Bias (SD) & ESE & CP95 & ESE.b & CP95.b  \\  \hline  
			 &   \multicolumn{15}{c}{ 	\multirow{2}{*}{IPW Estimator ($\hat \tau_{ipw}$), with True Propensity Score} }    \\
	 &    \multicolumn{15}{c}{ }    \\
(13) & 1.7 (210.0) & 209.5 & 94.9 & 206.5 & 94.4 & 5.6 (152.6) & 148.8 & 94.3 & 147.8 & 93.2 & 5.9 (102.0) & 105.6 & 94.6 & 104.6 & 95.2  \\
(14) & -1.3 (253.1) & 238.7 & 92.3 & 234.5 & 91.7 & -7.7 (169.6) & 170.0 & 95.2 & 169.1 & 94.6 & -6.7 (124.6) & 121.4 & 94.4 & 121.9 & 94.3  \\
	\hline
	 &    \multicolumn{15}{c}{ 	\multirow{2}{*}{IPW Estimator ($\tilde \tau_{ipw}$), with Estimated Propensity Score} }    \\
	 &    \multicolumn{15}{c}{ }    \\
(13)  & 6.9 (69.6) & 72.4 & 96.0 & 105.5 & 98.5 & 5.0 (47.3) & 47.0 & 95.1 & 52.5 & 95.9 & 4.8 (33.6) & 33.0 & 94.5 & 34.9 & 95.0   \\
(14) & -3.1 (140.5) & 125.2 & 91.0 & 162.7 & 95.4 & -1.3 (92.6) & 88.9 & 93.9 & 97.6 & 94.9 & -4.7 (69.9) & 63.7 & 92.8 & 66.9 & 94.0  \\
	\hline			
	 &    \multicolumn{15}{c}{ 	\multirow{2}{*}{Doubly Robust Estimator ($\hat \tau_{dr}$)} }    \\
	 &    \multicolumn{15}{c}{ }    \\
(13) & -2.6 (59.5) & 66.0 & 97.0 & 57.8 & 94.1 & 0.2 (45.8) & 48.7 & 96.6 & 44.2 & 94.8 & 3.0 (33.5) & 37.6 & 97.1 & 34.1 & 94.6  \\
(14) & -10 (118.6) & 119.1 & 94.3 & 105.1 & 91.0 & -4.4 (84.9) & 87.7 & 95.0 & 81.3 & 92.5 & -6.0 (67.8) & 65.0 & 94.1 & 62.2 & 92.6  \\
  \hline
 		 &   \multicolumn{15}{c}{ 	\multirow{2}{*}{\cite{athey2016estimating}'s Method} }    \\
	 &    \multicolumn{15}{c}{ }    \\
(13) & -46.0 (211.8) & 214.0 & 94.8 & 208.2 & 93.9 & -41.9 (153.7) & 151.1 & 92.9 & 149.6 & 91.9 & -44.3 (106.7) & 109.2 & 94.7 & 110.7 & 93.2 \\
(14) &-104.3 (256.1) & 244.2 & 91.9 & 237.0 & 89.8 & -114.2 (171.7) & 172.7 & 89.8 & 170.6 & 88.1 & -118.2 (124.0) & 122.9 & 83.1 & 122.4 & 81.9   \\
	\hline
\end{tabular}}
\begin{flushleft}
{\footnotesize Note: All the values in this table have been magnified 100 times. Bias and SD are the Monte Carlo bias and   
standard deviation over the 1000 simulations of the points estimates. 
 ESE and CP95 are the averages of estimated asymptotic standard error and coverage proportions of the 95\% confidence intervals based on the plug-in method, respectively. ESE.b and CP95.b have the same meaning as ESE and CP95 but are derived from 200 bootstraps.}
\end{flushleft}
\label{tab-s1}
\end{table}

\begin{table}
\centering
\caption{Comparison of various estimators for cases (15)-(16),  binary outcome.}  
\resizebox{\linewidth}{!}{\begin{tabular}{c rcccc  rcccc rcccc } 
\hline 
	 &    \multicolumn{5}{c}{$n_1 =50$}    & \multicolumn{5}{c}{$n_1 =100$} &  \multicolumn{5}{c}{$n_1 =200$}  \\
Case    & Bias (SD) & ESE & CP95 & ESE.b & CP95.b  & Bias (SD) & ESE & CP95 & ESE.b & CP95.b    & Bias (SD) & ESE & CP95 & ESE.b & CP95.b  \\  \hline  
			 &   \multicolumn{15}{c}{ 	\multirow{2}{*}{IPW Estimator ($\hat \tau_{ipw}$), with True Propensity Score} }    \\
	 &    \multicolumn{15}{c}{ }    \\
(15) & -0.2 (20.5) & 20.2 & 93.6 & 20.0 & 93.1 & 0.3 (14.2) & 14.6 & 94.9 & 14.6 & 94.7 & -0.2 (11.0) & 10.6 & 94.8 & 10.6 & 94.3    \\
(16) &  0.1 (25.9) & 25.7 & 93.7 & 25.5 & 93.6 & 1.0 (18.2) & 18.6 & 95.1 & 18.6 & 94.4 & -0.2 (14.1) & 13.6 & 94.4 & 13.8 & 94.3 \\
	\hline
	 &    \multicolumn{15}{c}{ 	\multirow{2}{*}{IPW Estimator ($\tilde \tau_{ipw}$), with Estimated Propensity Score} }    \\
	 &    \multicolumn{15}{c}{ }    \\
(15) &  0.1 (8.2) & 8.3 & 95.3 & 10.0 & 96.9 & 0.2 (6.0) & 6.2 & 95.5 & 6.5 & 95.7 & -0.1 (5.1) & 5.0 & 94.6 & 5.2 & 95.1   \\
(16) & 0.3 (9.4) & 8.9 & 93.7 & 12.3 & 97.4 & 0.2 (7.2) & 7.0 & 93.4 & 7.7 & 95.6 & 0.0 (6.7) & 6.1 & 90.7 & 6.6 & 92.5    \\
	\hline			
	 &    \multicolumn{15}{c}{ 	\multirow{2}{*}{Doubly Robust Estimator ($\hat \tau_{dr}$)} }    \\
	 &    \multicolumn{15}{c}{ }    \\
(15) &  -0.3 (8.2) & 7.3 & 92.7 & 8.5 & 95.2 & 0.1 (6.2) & 5.4 & 91.5 & 6.7 & 96.0 & -0.1 (5.3) & 4.1 & 86.5 & 5.6 & 95.3    \\
(16) & -0.8 (9.5) & 8.3 & 91.9 & 9.8 & 94.8 & -1.0 (8.4) & 6.9 & 89.5 & 8.4 & 94.6 & -1.0 (7.6) & 6.0 & 87.8 & 7.6 & 92.1   \\  
  \hline
 		 &   \multicolumn{15}{c}{ 	\multirow{2}{*}{\cite{athey2016estimating}'s Method} }    \\
	 &    \multicolumn{15}{c}{ }    \\
(15) &  -2.6 (20.2) & 20.1 & 93.8 & 19.6 & 93.0 & -2.1 (13.9) & 14.2 & 95.3 & 14.2 & 94.3 & -2.5 (10.5) & 10.0 & 93.2 & 10.0 & 93.2 \\
(16) & -6.3 (24.8) & 25.0 & 94.8 & 24.6 & 93.4 & -5.3 (17.6) & 17.6 & 94.5 & 17.5 & 92.8 & -6.2 (12.8) & 12.6 & 91.2 & 12.5 & 90.3   \\
	\hline
\end{tabular}}
\begin{flushleft}  
{\footnotesize Note: All the values in this table have been magnified 100 times. Bias and SD are the Monte Carlo bias and   
standard deviation over the 1000 simulations of the points estimates. 
 ESE and CP95 are the averages of estimated asymptotic standard error and coverage proportions of the 95\% confidence intervals based on the plug-in method, respectively. ESE.b and CP95.b have the same meaning as ESE and CP95 but are derived from 200 bootstraps.}
\end{flushleft}
\label{tab-s2}
\end{table}

\section{Additional Results for Application}
We report the real-data analysis results where the asymptotic standard errors are computed with the plug-in method, which are presented in Table \ref{tab-s3}.

\begin{table} 
\centering
\caption{Estimated effects of HCQ on renal failure.} 
\resizebox{\linewidth}{!}{\begin{tabular}{cc cc  cc cc } 
\hline
	 &   & \multicolumn{2}{c}{proportion = 0.3}    & \multicolumn{2}{c}{proportion = 0.4} &  \multicolumn{2}{c}{proportion = 0.5}  \\
End time  &   & Endpoint 1 & Endpoint 2 &  Endpoint 1 & Endpoint 2 & Endpoint 1 & Endpoint 2 \\  \hline  
			 &  &   \multicolumn{6}{c}{ 	\multirow{2}{*}{IPW, with True Propensity Score} }    \\
	 &  &  \multicolumn{6}{c}{ }    \\
 \multirow{2}{*}{3} &  Estimate (ESE) & -0.376 (0.100) & -0.202 (0.090) & -0.162 (0.062) & -0.202 (0.090) & -0.123 (0.068) & -0.202 (0.090)     \\
  & $p$-value  &  $<10^{-3}$ & 0.013  & 0.005  & 0.013  & 0.035  & 0.013     \\
 \multirow{2}{*}{4} & Estimate (ESE) & -0.418 (0.107) & -0.157 (0.076) & -0.211 (0.070) & -0.157 (0.076) & -0.172 (0.074) & -0.157 (0.076)  \\
  & $p$-value  &   $<10^{-3}$ & 0.020  & 0.002 & 0.020  & 0.010  & 0.020  \\ 
 \multirow{2}{*}{5} &Estimate (ESE) & -0.488 (0.113) & -0.192 (0.082) & -0.265 (0.075) & -0.192 (0.082) & -0.211 (0.077) & -0.192 (0.082)  \\
  & $p$-value  & $<10^{-3}$ & 0.010  & $<10^{-3}$ & 0.010  & 0.003& 0.010  \\ 
	\hline
	 &  &  \multicolumn{6}{c}{ 	\multirow{2}{*}{IPW, with Estimated Propensity Score} }    \\
	 &  &  \multicolumn{6}{c}{ }    \\
 \multirow{2}{*}{3} &  Estimate (ESE) & -0.318 (0.063) & -0.178 (0.074) & -0.138 (0.051) & -0.178 (0.074) & -0.102 (0.062) & -0.178 (0.074)   \\
  & $p$-value  &  $<10^{-3}$ & 0.008 & 0.004  & 0.008 & 0.051  & 0.008     \\
 \multirow{2}{*}{4} & Estimate (ESE) & -0.360 (0.063) & -0.145 (0.065) & -0.182 (0.055) & -0.145 (0.065) & -0.155 (0.063) & -0.145 (0.065)   \\
  & $p$-value  &  $<10^{-3}$  & 0.013 & $<10^{-3}$  & 0.013 & 0.007 & 0.013   \\ 
 \multirow{2}{*}{5} &Estimate (ESE) & -0.423 (0.062) & -0.178 (0.070) & -0.232 (0.055) & -0.178 (0.070) & -0.189 (0.064) & -0.178 (0.070)   \\
  & $p$-value  & $<10^{-3}$ & 0.006  & $<10^{-3}$ & 0.006  & 0.002  & 0.006   \\ 
	\hline			
  &  &   \multicolumn{6}{c}{ 	\multirow{2}{*}{Doubly Robust Estimator} }    \\
	 &  &  \multicolumn{6}{c}{ }    \\ 
 \multirow{2}{*}{3} &  Estimate (ESE) &  -0.256 (0.086) & -0.187 (0.063) & -0.073 (0.077) & -0.187 (0.063) & -0.034 (0.053) & -0.187 (0.063)  \\
  & $p$-value  &   0.001 & 0.001 & 0.173  & 0.001& 0.258  & 0.001    \\
 \multirow{2}{*}{4} & Estimate (ESE) & -0.337 (0.088) & -0.148 (0.050) & -0.125 (0.085) & -0.148 (0.050) & -0.092 (0.059) & -0.148 (0.050)  \\
  & $p$-value  &  $<10^{-3}$ & 0.002  & 0.071 & 0.002  & 0.058 & 0.002  \\ 
 \multirow{2}{*}{5} &Estimate (ESE) &  -0.396 (0.091) & -0.195 (0.057) & -0.169 (0.093) & -0.195 (0.057) & -0.119 (0.064) & -0.195 (0.057)\\
  & $p$-value  & $<10^{-3}$ & $<10^{-3}$ & 0.035  & $<10^{-3}$ & 0.032 & $<10^{-3}$  \\ 
 \hline   
 		 & &   \multicolumn{6}{c}{ 	\multirow{2}{*}{\cite{athey2016estimating}'s Method} }    \\
 	 &  &  \multicolumn{6}{c}{ }    \\
 \multirow{2}{*}{3} &  Estimate (ESE) & -0.034 (0.138) & -0.050 (0.070) & -0.045 (0.075) & -0.050 (0.070) & -0.031 (0.053) & -0.050 (0.070)      \\
  & $p$-value  &   0.402 & 0.239 & 0.275  & 0.239 & 0.282  & 0.239   \\
 \multirow{2}{*}{4} & Estimate (ESE) & -0.039 (0.150) & -0.043 (0.063) & -0.071 (0.086) & -0.043 (0.063) & -0.046 (0.064) & -0.043 (0.063)  \\
  & $p$-value  &  0.397 & 0.247  & 0.205 & 0.247  & 0.235 & 0.247   \\ 
 \multirow{2}{*}{5} &Estimate (ESE) & -0.053 (0.159) & -0.023 (0.071) & -0.082 (0.099) & -0.023 (0.071) & -0.060 (0.075) & -0.023 (0.071) \\
  & $p$-value  & 0.371 & 0.370 & 0.203  & 0.370 & 0.210 & 0.370   \\ 
  \hline   
%		 &  &  \multicolumn{6}{c}{ 	\multirow{2}{*}{Efficient Doubly Robust Estimator, with Estimated Propensity Score} }    \\
%	 &  &  \multicolumn{6}{c}{ }    \\
% \multirow{2}{*}{3} & Estimate (ESE) &  -0.062 (0.253) & -0.187 (0.06) & -0.312 (0.263) & -0.187 (0.06) & -0.674 (0.68) & -0.187 (0.06)  \\
%  & $p$-value  &     \\
% \multirow{2}{*}{4} & Estimate (ESE) & -0.162 (0.209) & -0.13 (0.054) & -0.353 (0.266) & -0.13 (0.054) & -0.713 (0.68) & -0.13 (0.054)\\
%  & $p$-value  &   \\ 
% \multirow{2}{*}{5} &Estimate (ESE) & -0.221 (0.209) & -0.175 (0.057) & -0.384 (0.269) & -0.175 (0.057) & -0.741 (0.682) & -0.175 (0.057)  \\
%  & $p$-value  &   \\ 
  \hline 
\end{tabular}}
\begin{flushleft}
{\footnotesize Note: ESE is estimated asymptotic standard 
error based on the plug-in method. The $p$-values are obtained by two-sided test, that is $H_{0}: \tau = 0$ against $H_{1}: \tau \neq 0$}
\end{flushleft}
\label{tab-s3}
\end{table}

\end{document}